\documentclass[letterpaper,reprint,aps,superscriptaddress,prl]{revtex4-1}
\usepackage[T1]{fontenc}
\usepackage{amsmath}
\usepackage{amssymb}
\usepackage{graphicx}
\usepackage{upgreek}

\usepackage[
colorlinks,
linkcolor=blue,
anchorcolor=black,
citecolor=blue,
urlcolor=blue,
]{hyperref}

\begin{document}
\title{Super-resolved imaging of a single cold atom on a nanosecond timescale}
\author{Zhong-Hua Qian}
\affiliation{CAS Key Laboratory of Quantum Information, University of Science and
Technology of China, Hefei 230026, China}
\affiliation{CAS Center For Excellence in Quantum Information and Quantum Physics,
University of Science and Technology of China, Hefei 230026, China}
\author{Jin-Ming Cui}
\email{jmcui@ustc.edu.cn}
\affiliation{CAS Key Laboratory of Quantum Information, University of Science and
Technology of China, Hefei 230026, China}
\affiliation{CAS Center For Excellence in Quantum Information and Quantum Physics,
University of Science and Technology of China, Hefei 230026, China}
\author{Xi-Wang Luo}
\affiliation{Department of Physics, The University of Texas at Dallas, Richardson,
Texas 75080-3021, USA}
\author{Yong-Xiang Zheng}
\affiliation{CAS Key Laboratory of Quantum Information, University of Science and
Technology of China, Hefei 230026, China}
\affiliation{CAS Center For Excellence in Quantum Information and Quantum Physics,
University of Science and Technology of China, Hefei 230026, China}
\author{Yun-Feng Huang}
\email{hyf@ustc.edu.cn}

\affiliation{CAS Key Laboratory of Quantum Information, University of Science and
Technology of China, Hefei 230026, China}
\affiliation{CAS Center For Excellence in Quantum Information and Quantum Physics,
University of Science and Technology of China, Hefei 230026, China}
\author{Ming-Zhong Ai}
\affiliation{CAS Key Laboratory of Quantum Information, University of Science and
Technology of China, Hefei 230026, China}
\affiliation{CAS Center For Excellence in Quantum Information and Quantum Physics,
University of Science and Technology of China, Hefei 230026, China}
\author{Ran He}
\affiliation{CAS Key Laboratory of Quantum Information, University of Science and
Technology of China, Hefei 230026, China}
\affiliation{CAS Center For Excellence in Quantum Information and Quantum Physics,
University of Science and Technology of China, Hefei 230026, China}
\author{Chuan-Feng Li}
\email{cfli@ustc.edu.cn}

\affiliation{CAS Key Laboratory of Quantum Information, University of Science and
Technology of China, Hefei 230026, China}
\affiliation{CAS Center For Excellence in Quantum Information and Quantum Physics,
University of Science and Technology of China, Hefei 230026, China}
\author{Guang-Can Guo}
\affiliation{CAS Key Laboratory of Quantum Information, University of Science and
Technology of China, Hefei 230026, China}
\affiliation{CAS Center For Excellence in Quantum Information and Quantum Physics,
University of Science and Technology of China, Hefei 230026, China}

\date{\today}

\begin{abstract}
	In cold atomic systems, fast and high-resolution microscopy of individual
	atoms is crucial, since it can provide direct information on the dynamics
	and correlations of the system. Here, we demonstrate nanosecond-scale two-dimensional
	stroboscopic pictures of a single trapped ion beyond the optical diffraction
	limit, by combining the main idea of ground-state depletion microscopy
	with  quantum state transition control in cold atoms. We achieve
	a spatial resolution up to 175~nm using an NA = 0.1 objective in the experiment, which represents a more than tenfold
	improvement compared with direct fluorescence imaging.  To
	show the potential of this method, we apply it to  observe the secular motion of the trapped ion, 
	we demonstrate a temporal resolution up to 50~ns with a displacement detection sensitivity of 10~nm. Our method
	provides a powerful tool for probing particle positions, momenta, 
	and correlations, as well as their dynamics in cold atomic systems. 
\end{abstract}
\maketitle

\emph{Introduction.---}Cold atomic systems, including cold quantum
gases in optical lattices, neutral atoms in optical tweezers, and trapped
ions, are promising platforms to study quantum simulation, computation,
and information processing. High-resolution optical detection and
imaging of individual particles in these systems are essential procedures, 
since they can provide direct information on quantum phenomena (e.g.,
transport, correlations, and phase transitions). In recent decades,
many microscope techniques have been developed for cold quantum gases
\citep{bakr2009aquantum,haller2015singleatom} and individual atoms
in optical tweezers \citep{kaufman2015entangling,levine2019parallel}
or ion traps \citep{nagerl1998coherent,wong-campos2016highresolution,zhukas2021directobservation}.
However, the resolution of these methods is fundamentally restricted to the wavelength
scale set by the optical diffraction limit, making them unsuitable
for probing quantum phenomena related to the details of the wave function 
in a variety of many-body systems.

Meanwhile, optical super-resolved microscopy has developed to maturity as
a powerful tool in chemistry and biology \citep{leung2011reviewof}. 
It allows chemical reactions and biological processes to be viewed on a  nanometer scale \citep{Hamans_2021,Sunbul_2021}.
In recent years, this method has also been applied to quantum systems, such as to imaging of solid-spin systems  
\citep{rittweger2009stedmicroscopy,han2009threedimensional,Maurer_2010,kolesov2018superresolution} and
to tracking of single-ion positions \citep{wong-campos2016highresolution,biercuk2010ultrasensitive,blums2018asingleatom},
and it has been  developed further in combination with quantum techniques 
for specific systems \citep{yavuz2007nanoscale,gorshkov2008coherent,cui2013quantum,tenne2019superresolution,yang2018theoryof}.
Realization of  detection and manipulation below the subwavelength
scale in cold atomic systems would allow  the study of quantum
transport, correlation, and dynamical phenomena in unprecedented ways, and therefore much research effort has been focused on  this area, 
with substantial progress being made in recent years. Examples include the generation of nanoscale optical potentials \citep{lkacki2016nanoscale,wang2018darkstate,ge2020darkstate},
stroboscopic painting of optical potentials for atoms with subwavelength
resolution \citep{lacki2019stroboscopic,tsui2020realization}, and observation of
the wave function of ensembles of atoms in one-dimensional optical lattices
based on nonlinear atomic responses \citep{mcdonald2019superresolution,subhankar2019nanoscale}
and pulsed ion microscopy \citep{veit2021pulsedion}. However, two-dimensional
(2D) super-resolved detection of a single cold atom below the optical
diffraction limit has yet to be demonstrated.
\begin{figure}
	\includegraphics[width=8.5cm]{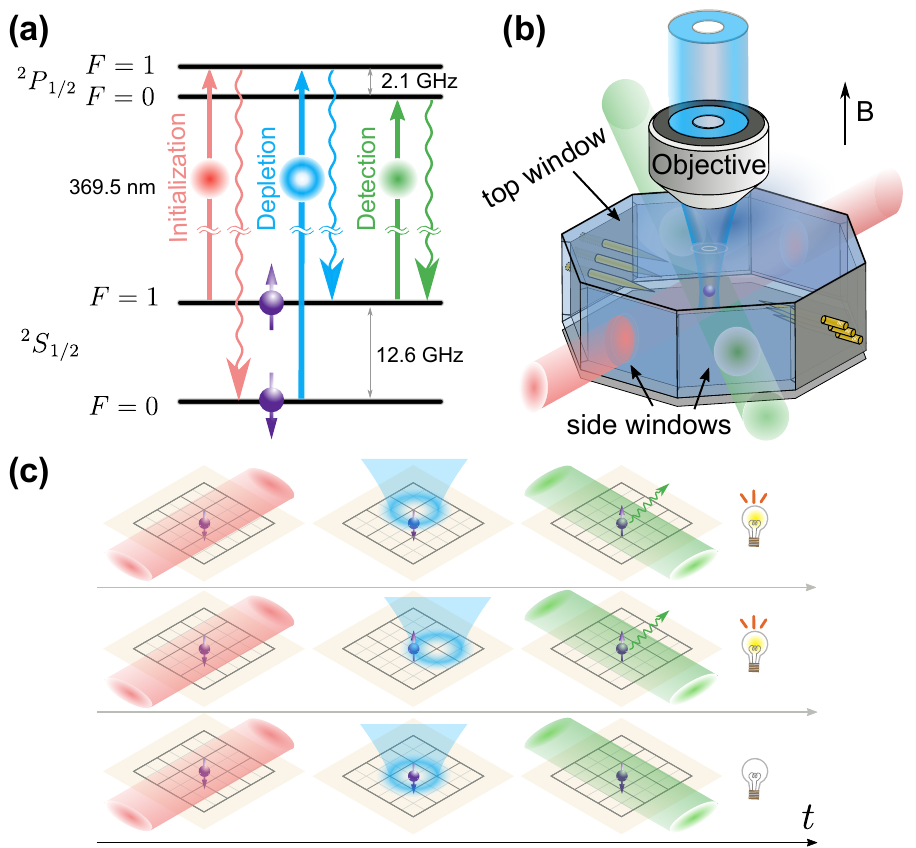}
	\caption{\label{fig:Fig1}Experimental scheme and setup. (a) Experimental
		scheme of energy levels of $^{171}{\rm Yb^{+}}$. The ion is optically
		pumped by a Gaussian beam (red line) and initialized to $^{2}\mathrm{S}_{1/2}|\mathrm{F}=0\rangle$
		(dark state). Depletion light (blue line) with a donut shape pumps the
		spin into $^{2}\mathrm{S}_{1/2}|\mathrm{F}=1\rangle$ (bright state).
		The detection laser (blue line) is used to read out the ion's state
		with state-dependent fluorescence. (b) Experimental setup: 
		a Paul trap in a vacuum chamber is used to trap the
		ion,  the initialization beam (red cylinder) and detection beam (green
		cylinder) are directed from the side windows of the vacuum chamber, and 
		the collimated donut beam is focused by the imaging objective (NA = 0.1).
		(c) Schematic of the super-resolved imaging process, in which three
		steps in (a) are performed to obtain a pixel of the image. A super-resolved image is obtained by scanning
		the donut spot, and only when the center of the donut spot is aligned
		with the ion does the ion avoid being polarized to the bright state and
	remains in the dark state.}
\end{figure}

Here we present a demonstration of ground-state depletion (GSD) microscopy
on a trapped ion system to achieve 2D imaging beyond the optical diffraction
limit. By combining the control sequence of the GSD microscopy with
the qubit states polarization and detection methods for the ion, a
2D image of a trapped ion with a resolution of 175~nm
is experimentally realized using an NA = 0.1 imaging objective, which
is a 13-fold improvement compared with direct fluorescence imaging.
Our microscopy method not only enables single-particle imaging of
cold atom systems to an unprecedented resolution, but, more importantly,
it also has the advantage of high temporal resolution. The depletion
process can be very short (around 50 ns) under strong laser power, which
enables us to take images in nanosecond intervals and thus allows
the measurement of atomic wave-function dynamics. To show the capability
of our method, we apply it to detect the secular motion of the trapped
ion. One cycle of the motion within 735 ns is observed, with a displacement
detection sensitivity of 10 nm.  Finally, we should point out that
our method is quite general and can be applied to both ions and
neutral atoms, and it can also be generalized to probe the many-body
position correlations and interacting dynamics of atomic arrays by
generating many independent depletion laser spots.

\emph{Experiment and results.---}The imaging process is based on
the transitions between ground states $^{2}\mathrm{S}_{1/2}$ and
first excitation states $^{2}\mathrm{P}_{1/2}$ of a $^{171}{\rm Yb^{+}}$
ion in a Paul trap (the transition wavelength is 369.5 nm), which
are sequentially controlled by different near-resonant lasers. The
energy level diagram of  $^{171}{\rm Yb^{+}}$ is presented in Fig.~\ref{fig:Fig1}(a). 
Three different lasers are primarily used
for the imaging process: the initialization laser resonates with $^{2}\mathrm{S}_{1/2}|\mathrm{F}=1\rangle\leftrightarrow {}^{2}\mathrm{P}_{1/2}|\mathrm{F}=1\rangle$,
which is used to polarize the nuclear spin states of the ion to $^{2}\mathrm{S}_{1/2}|\mathrm{F}=0\rangle$
(dark state); the depletion laser nearly resonates with $^{2}\mathrm{S}_{1/2}|\mathrm{F}=0\rangle\leftrightarrow {}^{2}\mathrm{P}_{1/2}|\mathrm{F}=1\rangle$,
which is used to depopulate the dark state to $^{2}\mathrm{S}_{1/2}|\mathrm{F}=1\rangle$
(bright state); the detection laser, with $^{2}\mathrm{S}_{1/2}|\mathrm{F}=1\rangle\leftrightarrow {}^{2}\mathrm{P}_{1/2}|\mathrm{F}=0\rangle$,
is used to detect nuclear spin in a dark or bright state. As
the transition between $^{2}\mathrm{S}_{1/2}|\mathrm{F}=0\rangle$
and $^{2}\mathrm{P}_{1/2}|\mathrm{F}=0\rangle$ is forbidden, the
state $\mathrm{S}_{1/2}|\mathrm{F}=1\rangle$ can continually scatter
fluorescence photons under the detection laser, which acts as a bright
state; by contrast, the state $\mathrm{S}_{1/2}|\mathrm{F}=0\rangle$
will not scatter detection photons, since it is not resonant with the
detection laser, and acts as a dark state \citep{Blinov_2004}. In
the depletion process, a donut-shaped beam is focused on the ion to polarize
its spin state through  spontaneous emission. The ion's
nuclear spin state is selectively polarized by scanning the donut
spot position, allowing the position information to be encoded into nuclear
spin states. Figure~\ref{fig:Fig1}(b) shows the configuration of
the three lasers in the experimental setup. The $^{171}{\rm Yb^{+}}$
ion is trapped in a Paul trap installed in a vacuum chamber \citep{Wang_2016}
, and a magnetic field of 9.7~G is applied along the $z$ axis.
The initialization laser and detection laser are directed to the trap
from the side windows of the chamber. These are Gaussian beams perpendicular
to the magnetic field with waists of 30 $\upmu$m. The depletion
light beam is  in a donut form and, along with the magnetic field, is focused
to a donut spot through an objective lens with NA = 0.1, and then shone
on the trapped ion across the top window of the vacuum chamber.

Figure~\ref{fig:Fig1}(c) illustrates the main control sequence for
the imaging process. The image is obtained pixel by pixel by scanning
the depletion spot position on a 2D plane and sequentially switching the 
three lasers on and off. To illustrate the basic principle of obtaining an image
of the ion, here we  take just three pixels as an example. Only when
the center of the donut spot is aligned with the ion does the ion avoid
being polarized to the bright state in the depletion process and remains in 
the dark state. To keep the ion at a low temperature during the imaging
process, we apply a 10~MHz red detuned laser containing frequencies
of $^{2}\mathrm{S}_{1/2}|\mathrm{F}=1\rangle\leftrightarrow {}^{2}\mathrm{P}_{1/2}|\mathrm{F}=0\rangle$
and $^{2}\mathrm{S}_{1/2}|\mathrm{F}=0\rangle\leftrightarrow {}^{2}\mathrm{P}_{1/2}|\mathrm{F}=1\rangle$
for Doppler cooling \citep{Olmschenk_2007}, and switch it on for 1~ms
cooling before each pixel imaging sequence \footnote{See Sec. II of the Supplemental Material at [URL will be inserted by publisher] for details of the laser system}.

\begin{figure*}
	\includegraphics[width=17cm]{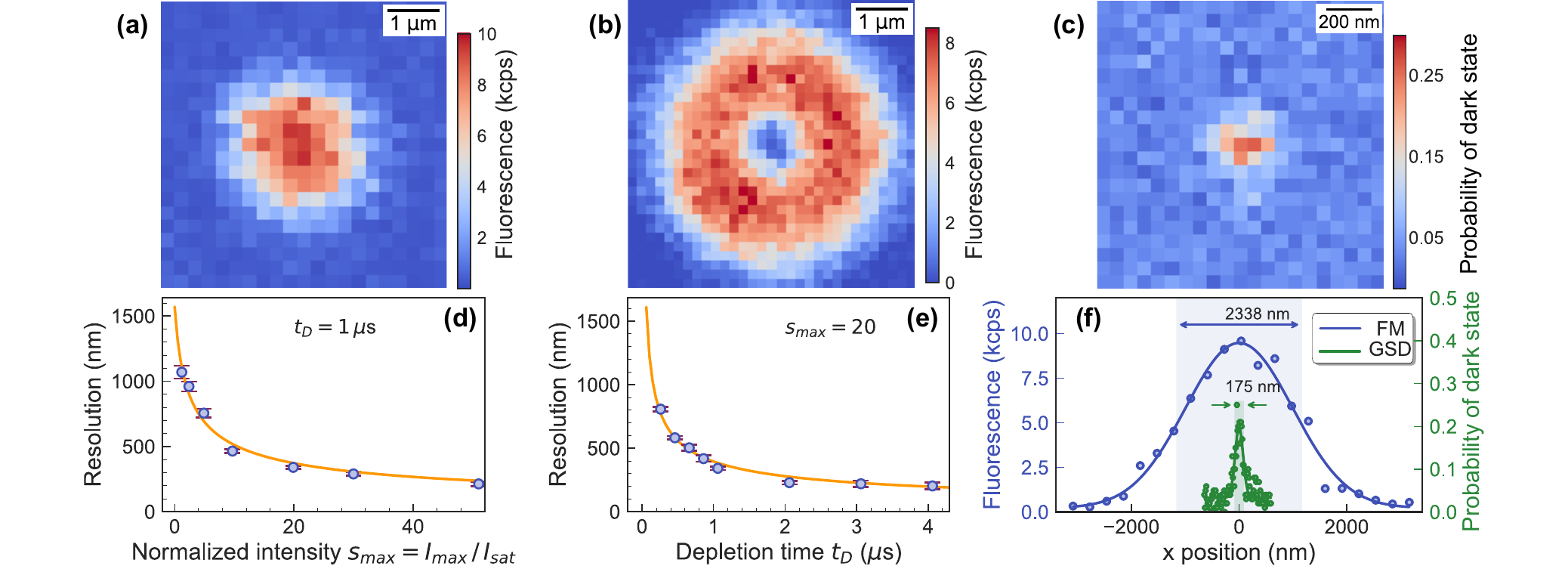}
	\caption{\label{fig:Fig2}Demonstration of super-resolved imaging of a trapped
		ion. (a) and (b) Use of the trapped ion as a probe to characterize the profile
		of Gaussian and donut spots, respectively, after aberration compensation using DMD with the holographic beam reshaping method \citep{zupancic2016ultraprecise}.
		The intensity is represented by the number of scattered fluorescence
		photons measured per second (kilocounts per second, kcps). The minimum
		intensity of the donut spot in the center is 3.8\% of the maximum
		intensity. (c) Super-resolved image of a single ion. The depletion
		laser is linearly polarized, and the probability of the dark state is obtained
		from the statistics of 100 measurements. (d) Spatial resolution (FWHM)
		as a function of normalized intensity $s_{\mathrm{max}}=I_{\mathrm{max}}/I_{\mathrm{sat}}$
		of donut light for 1~$\upmu$s depletion time. (e) Spatial
		resolution as a function of depletion time with $s_{\mathrm{max}}$
		fixed at 20. (f) Contrast of resolutions between fluorescence
		microscope and GSD microscope with depletion time $t_D=7\,\upmu s$ and depletion power $P=65$~nW ($s_{\text{max}}=14$). 
		The solid lines are fits with Gaussian functions.}
\end{figure*}

To realize high-spatial-resolution imaging in GSD, a highly focused
donut spot with high mode purity is crucial, requiring the dark center's
residual intensity to be as low as possible. To generate a donut spot with
high purity, we employ a holographic beam reshaping method with a
digital micromirror device (DMD) \citep{zupancic2016ultraprecise}.
Compared with other methods \citep{Beijersbergen_1994,Oemrawsingh_2004,Beijersbergen_1993,Petrov_1997,Slussarenko_2011,D_Ambrosio_2013,Yan_2015,Ye_Fang_Wei_2003,Qi_Xiao_Qing_2010},
we can measure the optical system aberration \emph{in situ} and then compensate for 
the aberration by this method. Our study first uses the trapped ion
as a probe to detect the interference light intensity and measure
the aberration phase map of the optical system in combination with the
DMD \footnote{See  Sec. III of the Supplemental Material at [URL will be inserted by publisher] for details of holographic beam reshaping with DMD.}. Subsequently, computer-generated holograph
patterns with aberration compensation are calculated and then programmed
on the DMD to generate the desired beam profiles. To characterize the
quality of focused spots after aberration compensation, we investigate
the profiles of reshaped Gaussian and donut spots [see Figs.~\ref{fig:Fig2}(a) and \ref{fig:Fig2}(b), 
respectively] by scanning the spots around the ion and detecting
the ion's fluorescence with a cooling laser. The Gaussian spot is
focused to a full width at half maximum (FWHM) of 2.34~$\upmu$m, which
is at its diffraction limit. For the donut beam, the residual light
intensity in the dark center is 3.8\% of the maximum intensity.

After preparation of the dark-center donut spot, the imaging sequences
in Fig.~\ref{fig:Fig1}(c) were applied with a linearly polarized donut
beam. A super-resolved image of a single trapped ion was obtained, 
as shown in Fig.~\ref{fig:Fig2}(c). The FWHM in the $x$ axis is 175~nm 
at 65~nW depletion laser power and 7~$\upmu$s
depletion time. A comparison of the measured
point spread functions (PSFs), shown in Fig.~\ref{fig:Fig2}(f), reveals that this technique gives a 13-fold improvement  in spatial resolution over
fluorescence microscopy. We
also analyzed the resolution as a function of the intensity and pulse
duration of the depletion light, and the results are shown in Figs.~\ref{fig:Fig2}(d) and \ref{fig:Fig2}(e). 
Here, for generality, the intensity is characterized by the normalized intensity $s_{\mathrm{max}}=I_{\mathrm{max}}/I_{\mathrm{sat}}$, 
where $I_{\mathrm{max}}$ denotes the maximum intensity
of the donut spot profile and $I_{\mathrm{sat}}=510$~W/m$^{2}$ is
the saturation intensity of the $\mathrm{Yb}^{+}$ ion. The results
are consistent with the spatial resolution $\Delta r$ of the simplified
model \footnote{See Sec. VIII of the Supplemental Material [URL will be inserted by publisher]}:
\begin{equation}
	\Delta r=\frac{\mathrm{FWHM}_{01}}{\sqrt{1+k s_{\mathrm{max}} t_{D}}}=\frac{\mathrm{FWHM}_{00}}{1.67\sqrt{1+k s_{\mathrm{max}} t_{D}}},\label{eq:resolution}
\end{equation}
where the coefficient $k=\beta\sigma I_{\mathrm{sat}}/\hbar\omega_{0}$,
$k s_{\mathrm{max}}$ represents the depletion rate, $\mathrm{FWHM}_{01}=1.41~\upmu$m
is the FWHM of the central hollow area of the donut spot ($\mathrm{LG}_{0}^{1}$
mode), $\mathrm{FWHM}_{00}=2.34~\upmu$m is the FWHM of the Gaussian
spot ($\mathrm{LG}_{0}^{0}$ mode), $\beta$ is
the branching ratio for the decay into the bright state, $\sigma$ is
the cross-section \citep{Foot2005}, and $t_{D}$ is the depletion
pulse duration. This model indicates that the resolution is determined
by the product of $s_{\mathrm{max}}$ and $t_{D}$. Theoretically, the
resolution can be extended to the subnanometer scale as long as we increase
$s_{\mathrm{max}}t_{D}$. However, further improvement in resolution
is limited by actual experiment conditions. First, the imperfection
of the donut spot results in residual light in the center. As the
donut power increases, the residual light in the center $s_{0}$ will 
be large enough to excite the spin and make the position information
indistinguishable from the probability noise. The resolution limit  \footnote{See  Sec. VIII of the Supplemental Material at [URL will be inserted by publisher] for a derivation of this limit.}
$\Delta r_{\mathrm{limit}}\propto \lambda/(\mathrm{NA}\sqrt{\mathrm{ER}})$, where $\mathrm{ER}=I_{\mathrm{max}}/I_{0}$
is defined as the extinction ratio of the doughnut spot. Second,
the spatial resolution is limited by the ion's wave-packet size after
cooling, which is about 20~nm, as given by the Doppler cooling limit in
the present study.

\begin{figure*}
	\includegraphics[width=17cm]{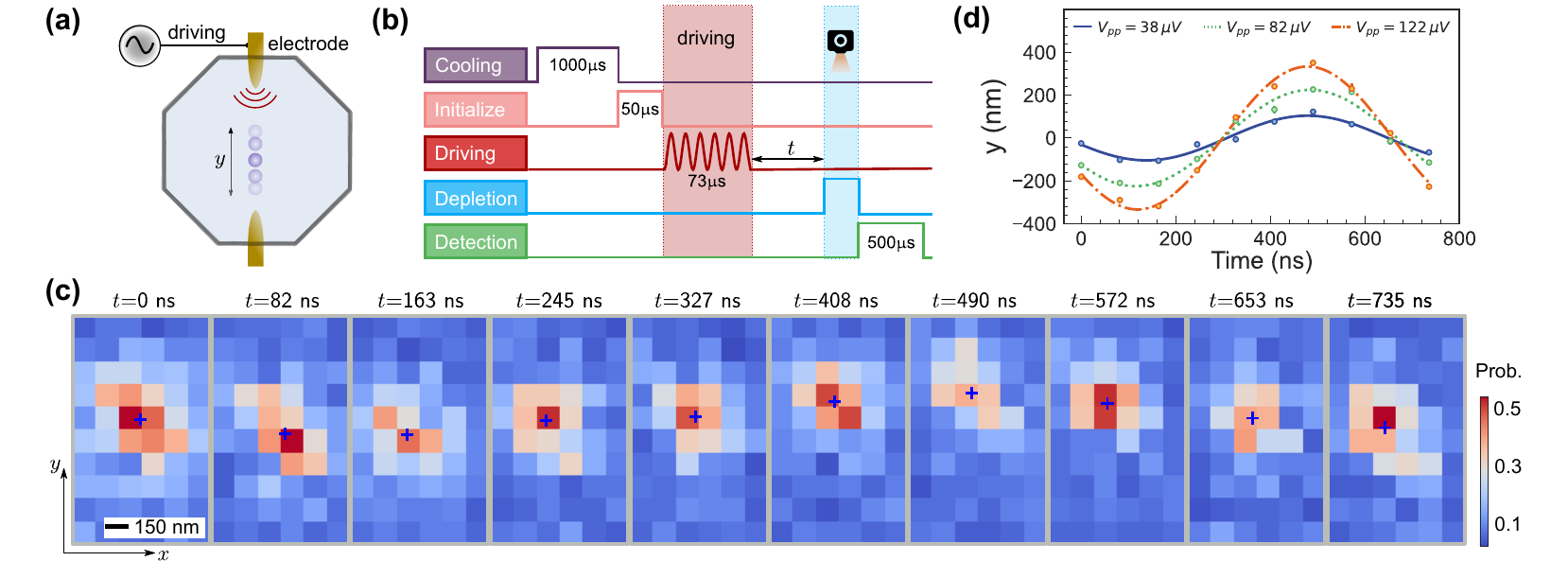}
	\caption{\label{fig:Fig3}Nanosecond photography to detect the dynamics of
		the trapped ion. (a) To excite the motion mode along the $y$ axis, a
		resonant electrical signal is applied to one electrode of the Paul
		trap, providing a sinusoidal driving force. (b) Sequence diagram of the
		imaging process. The ion is driven to steady oscillation by a resonant
		pulse, with a driving voltage  $V_{\mathrm{pp}}$ and a driving
		duration  of 73.5~$\upmu$s. The ion is then imaged at $t$ after 
		the driving has been stopped, with a 50~ns depletion duration. $s_{\mathrm{max}}$
		is fixed at 450. (c) Photographs of the moving ion driven by $V_{\mathrm{pp}}=38~\upmu$V at different 
		$t$. The blue crosses  mark the fitted centers. (d) The ion's  motional trajectories along the $y$ axis under different
	driving voltages, which are constructed by the fitted centers of super-resolved images.} 
\end{figure*}
Next, we will present a method to detect the dynamics of a single cold
atom in a 2D plane with high precision. Because the ion is trapped
in a harmonic trap, it undergoes secular oscillations in three axes \citep{leibfried2003quantum}. 
Here, we select the oscillation along  two RF needles (the $y$ axis)
for the study, as shown in Fig.~\ref{fig:Fig3}(a). To demonstrate this application, we first drive the
ion resonantly with an electrical pulse to excite the oscillation
along the $y$ axis after Doppler cooling. The electrical pulse provides
a driving force in the form  $F_{d}\sin(\omega_{d}t)$, which drives
at a frequency  $\omega_{d}$ and lasts for a time $t_{d}$. During
the driving process, the cooling lasers are turned off to avoid motion
dissipation. When the driving pulse ends, the ion is drived into a steady-state
oscillation, and the amplitude of the oscillation is given by 
\begin{equation}
	A=\frac{2F_{d}}{m(\omega_{y}^{2}-\omega_{d}^{2})}\sin\!\left[\frac{(\omega_{y}-\omega_{d})t_{d}}{2}\right],\label{eq:motion}
\end{equation}
where $\omega_{y}=2\pi\times1.36$~MHz is the oscillation frequency
of the ion, $t_d=100T=100\cdot 2\pi/\omega_y$, and $m$ is the ion mass. Near resonance $(\omega_{d}\approx\omega_{y})$,
we find that $A=F_{d}t_{d}/2m\omega_{y}$, which indicates that the amplitude of motion
is proportional to the driving force $F_{d}$ and driving
time $t_{d}$. 

To acquire the instantaneous position of the ion at time $t$ after
the driving pulse, we implement the super-resolved microscopy with
a delay time $t$ before the depletion process. To get a pixel in an image, a control sequence in Fig.~\ref{fig:Fig3}(b) is repeated 100 times to get the dark state probability, 
which requires 0.16~s. 
As the excited oscillation is synchronised to the driving wave phase, we fix the driving wave phase to make sure the current position of the ion is constant between different shots.
To obtain one image, the
donut beam is scanned to $10\times10$ pixels, while  the depletion
pulse duration is kept at 50~ns, which is much shorter than the period of motion 
(735~ns). The normalized intensity of the doughnut spot $s_{\mathrm{max}}$
is fixed at 450. The total time to get an image is 16~s. By imaging the ion at different $t$, we can see its dynamics
in the trap. Figure~\ref{fig:Fig3}(c) shows a typical
photographic recording of the oscillating ion in the central $6\times 10$ pixels. By fitting each image to
obtain the displacements of the ion at different $t$, the ion trajectory
can be reconstructed with a precision of 10~nm. The fitted spatial resolution is averaged as 373 nm. The ion's oscillating
trajectories under different driving forces are also investigated
in this study, as shown in Fig.~\ref{fig:Fig3}(d). The result shows that the amplitude of 
motion  depends linearly on the driving force, as predicted by
Eq.~\eqref{eq:motion}. We observed an amplitude of motion of $104.2\pm8.9$~nm under a 38~$\upmu$V peak-to-peak driving voltage ($V_{\mathrm{pp}}=38~\upmu$V).
The corresponding driving force $F_{d}=6.9 \pm 0.7$~zN. The period of motion is fitted as $681.1 \pm23.9$~ns, which has a
7\%  error relative to the calculated period. 

\emph{Outlook and conclusions.---} In this work, we have demonstrated a
nanosecond-scale super-resolved 2D imaging method for a single
trapped ion. Our approach is quite general and can be applied to
both charged atoms (e.g., the trapped ion studied here) and neutral ultracold
atoms (e.g., a single atom trapped in an optical tweezer), for which the
holographic beam reshaping method with DMD~\citep{zupancic2016ultraprecise}
as well as  (nuclear-) spin polarization and detection have already been
widely used. Moreover, an arbitrary array of spots with various shapes
can be easily generated at the same time, using the holographic beam
reshaping method \citep{zupancic2016ultraprecise}, and each of them
can be controlled independently. Therefore, by  applying
these depletion spots in parallel on site-resolved cold atoms (e.g., tweezer arrays),
we can develop this imaging method to probe  two- or multisite
correlations and their interacting many-body dynamics. 

It is worth noting that the resolution of the microscopy can be improved
by using higher-NA objectives under the same depletion parameters
(e.g., the NAs of objectives have reached 0.6 for trapped ion systems
\citep{wong-campos2016highresolution} and 0.7 for cold neutral atoms
in optical tweezers \citep{kaufman2015entangling}), and the GSD schemes
will support an imaging resolution below 30~nm and a displacement measurement
accuracy below 2~nm in principle, which is below the level of the wave-packet size of the motional ground state. Furthermore,
the limit of resolution  is approximately $\mathrm{FWHM}_{01}/\sqrt{-\mathrm{ER}\ln{\Delta p}}$
\footnote{See Sec. VIII of the Supplemental Material at [URL will be inserted by publisher] for a derivation of this limit.},
by reducing the ER \citep{zupancic2016ultraprecise,RN1361,RN1651} and improving the detection fidelity of the dark state
$\Delta p$ \citep{noek_high_2013}, the resolution can further reach 3.1~nm under an objective with NA=0.6.

Together with the advantage of nanosecond imaging time, our work opens
up new opportunities in the study of  single-particle dynamics,
many-body correlation~\citep{bergschneider2019experimental}, and
even  two-body collision dynamics~\citep{guan_density_2019} in
cold atomic systems. For example, long-range interactions based on
Rydberg atoms and ions could induce two-particle entanglement in position
and momentum even when they are far apart (with site addressability), and 
applying  fast and high-resolution imaging of each particle allows
direct measurement of both two-particle position correlation and
particle dynamics, with no need for any additional mapping or time-of-flight operation
\citep{bergschneider2019experimental}. Our method can
even be used to measure  short-range correlations with contact
interactions (e.g., the spin-exchange interaction \citep{kaufman2015entangling,bergschneider2019experimental}),
as long as the two interacting atoms are distinguishable particles
(e.g., different species or spins). We can first coherently transfer
a spin-up (or spin-down) state from the $S$ to the $D$ level to store the
spin, and then apply the GSD sequence to measure the position of the spin-down
(or spin-up) state, while  using different wavelengths to image different
species.

In conclusion, we have presented a nanosecond-scale super-resolved imaging method
to detect a single trapped ion and its dynamics in two dimensions,
achieving a spatial-time  resolution to 175~nm at 7~$\upmu$s and 373~nm at 50~ns, and the resolution can be improved
further using higher-NA objectives. The method is general and can
be used for both charged and neutral cold atoms. It can also be developed further  
to detect many-body correlations using multiple GSD spots. Furthermore,
the GSD method on a cold atom system can be extended in the future to three dimensions,
like classical GSD \citep{han2009threedimensional}, thereby further enhancing its potential.

\begin{acknowledgments}
	\emph{Note added.---}During preparation of the manuscript, we became aware
	of recent work on optical super-resolution sensing of a trapped
	ion's wave-packet size \citep{drechsler_optical_2021}, which uses the ground-state
	depletion technique to detect the size of the motion wave packet  of a trapped
	Ca$^{+}$ ion.

	This work was supported by the National Key Research and Development
	Program of China (Grant Nos. 2017YFA0304100 and 2016YFA0302700),
	the National Natural Science Foundation of China (Grants Nos. 11821404,
	11874343, 11774335, and  11734015), the Key Research Program
	of Frontier Sciences, CAS (Grant No. QYZDY-SSW-SLH003), the Science Foundation
	of the CAS (Grant No. ZDRW-XH-2019-1), the Fundamental Research Funds
	for the Central Universities (Grant Nos. WK2470000026, WK2470000027,
	and WK2470000028), the Anhui Initiative in Quantum Information Technologies
	(Grant Nos. AHY020100 and  AHY070000), and the National Program for
	Support of Topnotch Young Professionals (Grant No. BB2470000005). 

	Z.-H.Q. and J.-M.C. contributed equally to this work.
\end{acknowledgments}


%

\end{document}


\title{Supplemental Material: Super-resolved imaging of a single cold atom
on a nanosecond timescale}
\author{Zhong-Hua Qian}
\affiliation{CAS Key Laboratory of Quantum Information, University of Science and
Technology of China, Hefei 230026, China}
\affiliation{CAS Center For Excellence in Quantum Information and Quantum Physics,
University of Science and Technology of China, Hefei 230026, China}
\author{Jin-Ming Cui}
\email{jmcui@ustc.edu.cn}

\affiliation{CAS Key Laboratory of Quantum Information, University of Science and
Technology of China, Hefei 230026, China}
\affiliation{CAS Center For Excellence in Quantum Information and Quantum Physics,
University of Science and Technology of China, Hefei 230026, China}
\author{Xi-Wang Luo}
\affiliation{Department of Physics, The University of Texas at Dallas, Richardson,
Texas 75080-3021, USA}
\author{Yong-Xiang Zheng}
\affiliation{CAS Key Laboratory of Quantum Information, University of Science and
Technology of China, Hefei 230026, China}
\affiliation{CAS Center For Excellence in Quantum Information and Quantum Physics,
University of Science and Technology of China, Hefei 230026, China}
\author{Yun-Feng Huang}
\email{hyf@ustc.edu.cn}

\affiliation{CAS Key Laboratory of Quantum Information, University of Science and
Technology of China, Hefei 230026, China}
\affiliation{CAS Center For Excellence in Quantum Information and Quantum Physics,
University of Science and Technology of China, Hefei 230026, China}
\author{Ming-Zhong Ai}
\affiliation{CAS Key Laboratory of Quantum Information, University of Science and
Technology of China, Hefei 230026, China}
\affiliation{CAS Center For Excellence in Quantum Information and Quantum Physics,
University of Science and Technology of China, Hefei 230026, China}
\author{Ran He}
\affiliation{CAS Key Laboratory of Quantum Information, University of Science and
Technology of China, Hefei 230026, China}
\affiliation{CAS Center For Excellence in Quantum Information and Quantum Physics,
University of Science and Technology of China, Hefei 230026, China}
\author{Chuan-Feng Li}
\email{cfli@ustc.edu.cn}

\affiliation{CAS Key Laboratory of Quantum Information, University of Science and
Technology of China, Hefei 230026, China}
\affiliation{CAS Center For Excellence in Quantum Information and Quantum Physics,
University of Science and Technology of China, Hefei 230026, China}
\author{Guang-Can Guo}
\affiliation{CAS Key Laboratory of Quantum Information, University of Science and
Technology of China, Hefei 230026, China}
\affiliation{CAS Center For Excellence in Quantum Information and Quantum Physics,
University of Science and Technology of China, Hefei 230026, China}
\date{\today}
\maketitle

\section{Ion trap setup}

\begin{figure*}[h]
\includegraphics[width=9cm]{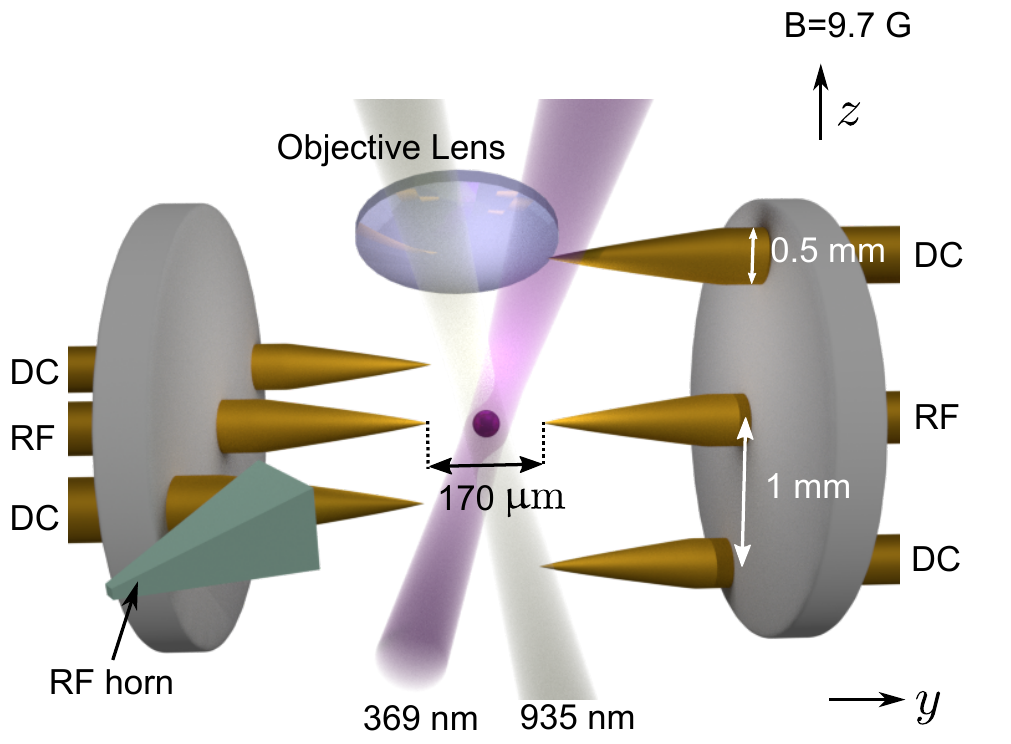} \caption{\label{fig:Fig1}Electrode configuration of the needle traps (not
to scale). }
\end{figure*}

The ion trap in this experiment is a three-dimensional (3D) needle
trap, as shown in Fig.~\ref{fig:Fig1}. The trap electrodes consist
of six tungsten needles \citep{Wang_2016}. Each needle has a diameter
of 0.5~mm, and adjacent needles are 1~mm apart. Opposite rows of
electrodes are perpendicular to each other. The two central electrodes
have a tip-to-tip distance of 170~$\upmu$m and are supplied with
22~MHz radio frequency (RF). All electrodes are supplied with a static
(DC) voltage ranging from $-5$~V to 5 V to modify the ion's position.
This 3D trap is suitable for trapping a single $^{171}\mathrm{Yb}^{+}$
ion. To provide a quantum axis for the spin, a 9.7~G magnetic field
along the $z$ axis is needed.

\section{Laser configuration for $^{171}\text{Yb}^{+}$ }

The state operation on $^{171}{\rm Yb}^{+}$ involves Doppler cooling,
pumping (or initialization), detection, and depletion. The $^{171}{\rm Yb}^{+}$
ion is Doppler-cooled by laser light whose frequency is $\Delta_{1}=10$~MHz
red-detuned from the transition $^{2}{\rm S_{1/2}|F=1\rangle}$ (bright
state) to $^{2}{\rm P_{1/2}|F=0\rangle}$ and $^{2}{\rm S_{1/2}|F=0\rangle}$
(dark state) to $^{2}{\rm P_{1/2}|F=1\rangle}$ \citep{Olmschenk_2007}.
A repumping beam (935~nm) is used to repump the ion from $^{2}\mathrm{D}_{3/2}$
to the $^{3}\mathrm{D}[3/2]_{1/2}$ state, from which the electron
will decay to the $^{2}\mathrm{S}_{1/2}$ state, forming a closed
cooling cycle. A 638~nm laser beam pumps the ion from $^{2}\mathrm{F}_{7/2}$
to $^{1}\mathrm{D}[5/2]_{5/2}$ so that it will decay back to the
cooling cycle. Initialization to the ground state is realized with
pumping light that is $\Delta_{2}=3$~MHz red-detuned from the transition
$^{2}{\rm S_{1/2}|F=0\rangle}$ to $^{2}{\rm P_{1/2}|F=1\rangle}$.
The ion is initialized into the dark state $^{2}{\rm S_{1/2}|F=0\rangle}$
within tens of microseconds. Detection is achieved by a state-dependent
fluorescence technique using laser light of the transition $^{2}{\rm S_{1/2}|F=1\rangle}$
to $^{2}{\rm P_{1/2}|F=0\rangle}$, which is also 3~MHz red-detuned
from resonance. The ion will scatter fluorescence photons only when
in the $^{2}{\rm S_{1/2}|F=1\rangle}$ state (bright state). The ion
is depopulated from the dark state into the bright state by the depletion
laser, whose wavelength is near-resonant with the transition between
$^{2}\mathrm{S}_{1/2}|\mathrm{F}=0\rangle$ and $^{2}\mathrm{P}_{1/2}|\mathrm{F}=1\rangle$.

\begin{figure}
\includegraphics[width=9cm]{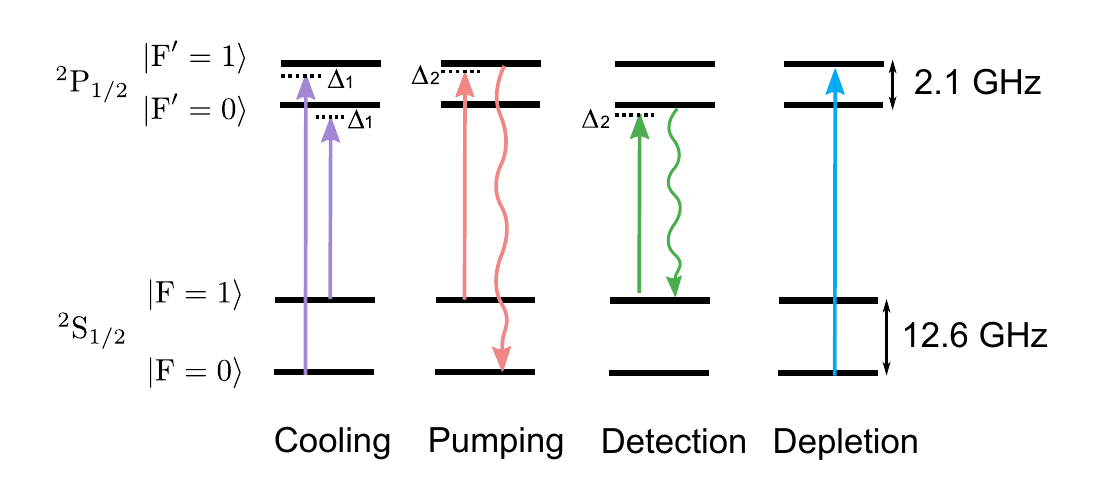} \caption{\label{fig:Fig2}$^{171}\text{Yb}^{+}$ energy level diagram for Doppler
cooling, pumping, detecting, and depletion.}
\end{figure}

\begin{figure}
\centering{}\includegraphics[width=12cm]{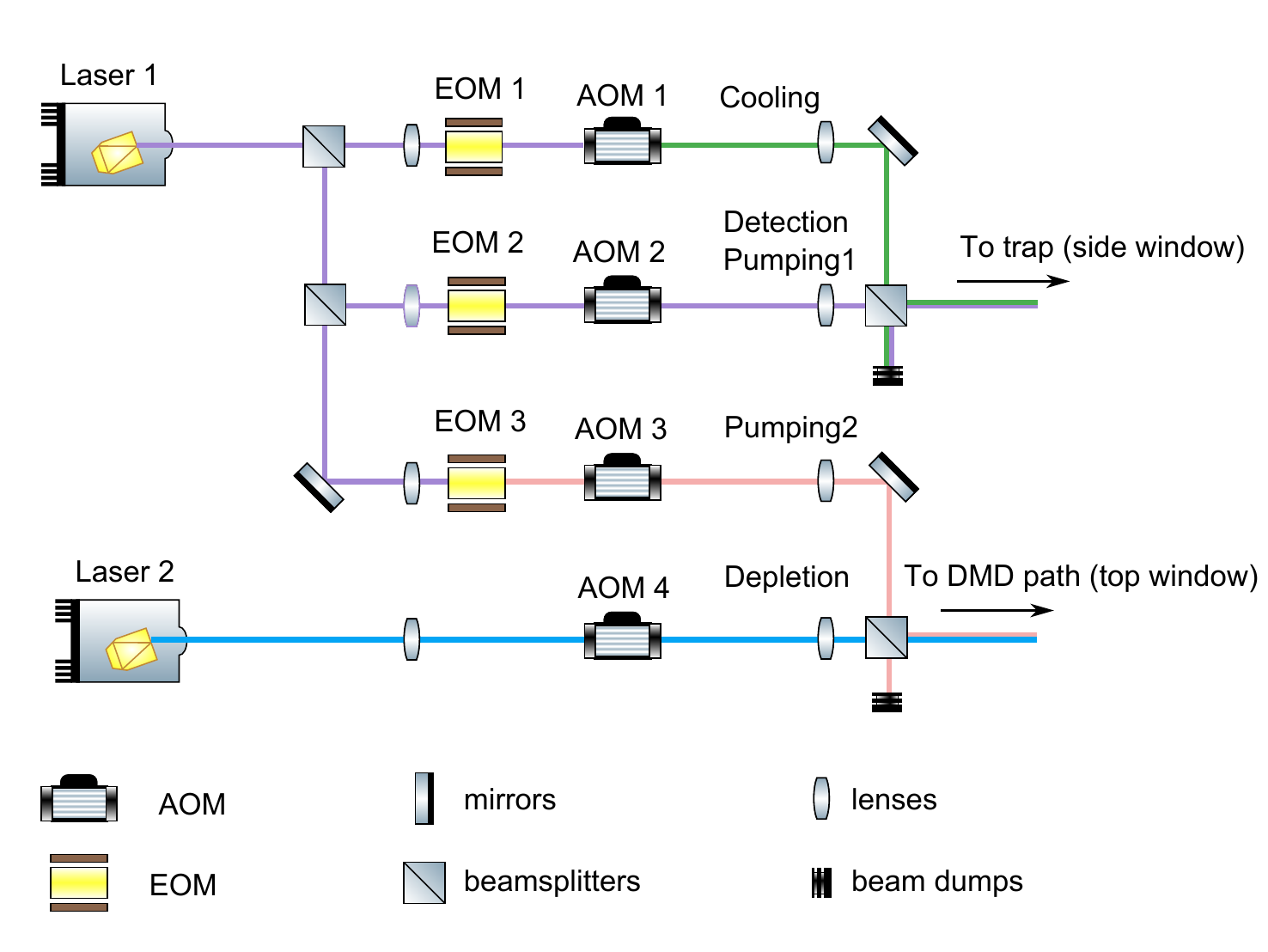} \caption{\label{fig:Fig3}Laser configuration for Doppler cooling, pumping,
detecting, and depletion.}
\end{figure}

\begin{figure}
\includegraphics[width=8cm]{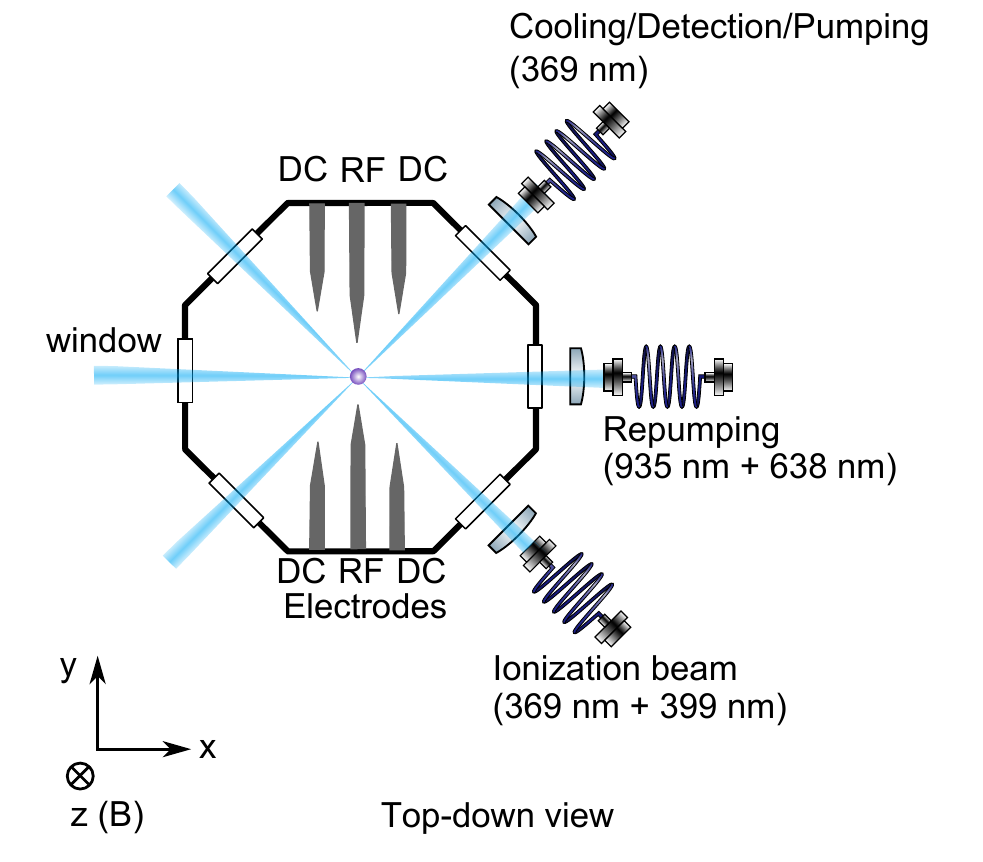} \caption{\label{fig:Fig4}Top view of the laser configuration of the 3D needle
trap. The waists of the cooling, repumping, and ionization beams are
20~$\upmu$m, 60~$\upmu$m, and 30~$\upmu$m, respectively.}
\end{figure}

\begin{figure}
\includegraphics[width=15cm]{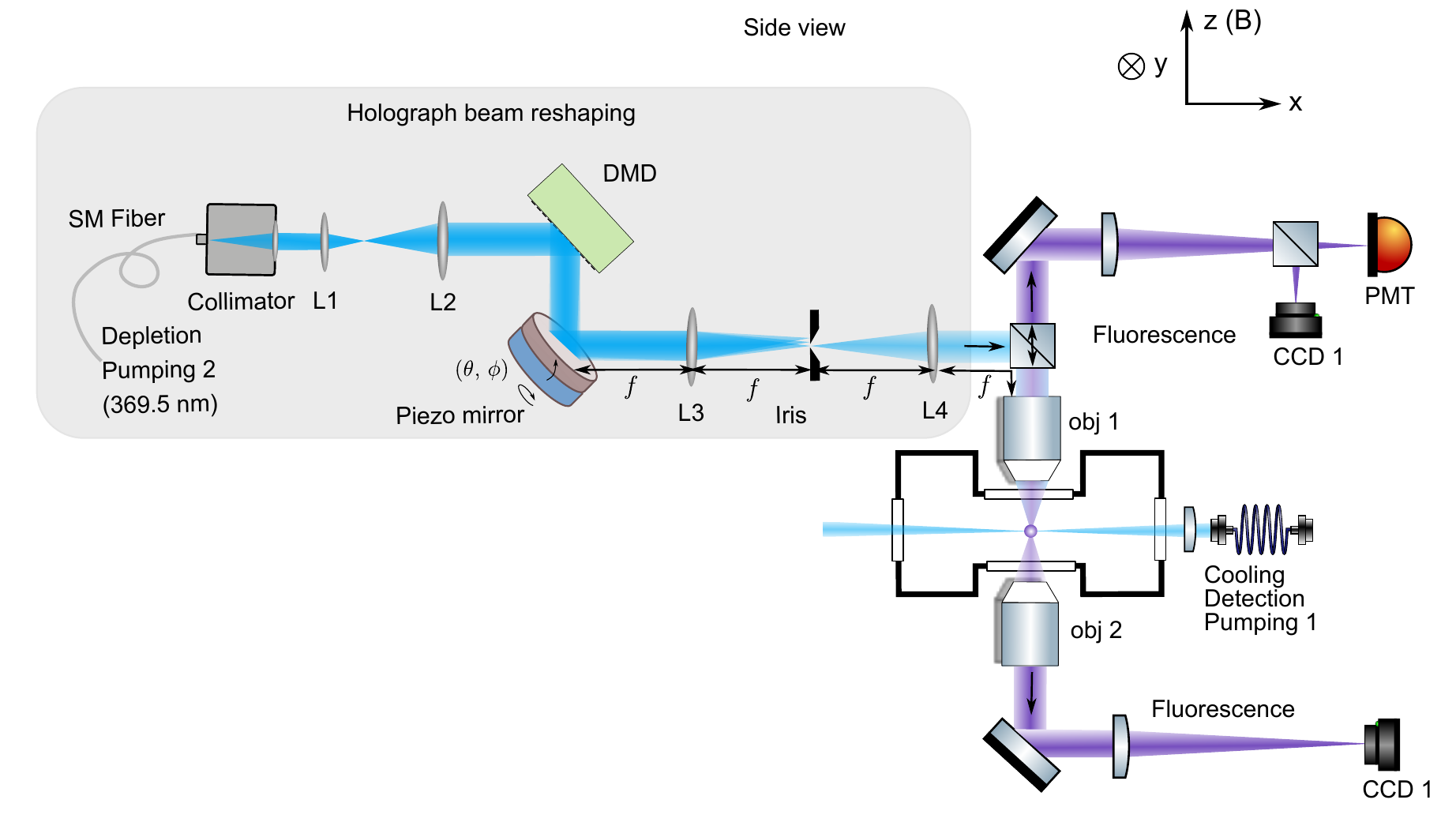} \caption{\label{fig:Fig5}Side view of the laser configuration of the needle
trap. The setup in the gray box is used to reshape the Gaussian beam
into a donut beam. The first-order diffraction beam is phase-modulated
and filtered by an iris. The objective obj 1 serves for state detection
and focusing of the donut beam, while obj 2 is used to precisely calibrate
the position of the donut spot.}
\end{figure}

The laser configuration is shown in Fig.~\ref{fig:Fig3}. Laser 1
(wavelength 369.52624~nm) provides cooling, pumping, and detection
beams. For the cooling laser, an electro-optic modulator (EOM 1) provides
a 14.7~GHz sideband. The detection beam and pumping beam share the
same path and are switched by turning off/on EOM 2 (2.1~GHz). These
two beams are combined and directed to the ion trap from the side
window (see Fig.~\ref{fig:Fig4}). The depletion beam is directed
to the digital micromirror device (DMD) path and is then is reshaped
into a donut beam (for the setup, see Fig.~\ref{fig:Fig5}), which
will be described later. Laser 2 (wavelength 369.51945~nm) is 14.7~GHz
blue-detuned with laser 1. Both lasers are frequency-stabilized within
1~MHz through the PDH locking technique \citep{Black_2001} by locking
to an ultrastable cavity. Acousto-optic modulators (AOMs) are used
to switch beams.

Of the laser beams directed into the side windows (Fig.~\ref{fig:Fig4}),
the Doppler cooling, detection, and pumping lasers are at $45^{\circ}$
to the $y$ axis to ensure that the $k$ vector is not orthogonal
to any motion mode. Thus, the ion will be efficiently Doppler-cooled
in three axes. The Yb atoms are ionized by a two-photon transition
with 369~nm and 399~nm ionization beams with powers of 200~$\upmu$W
and 500~$\upmu$W, respectively.

The depletion beam is reshaped into a donut beam and is then incident
into objective 1 (NA = 0.1). The beam shaping setup is illustrated
in Fig.~\ref{fig:Fig5}. The objectives obj 1 and obj 2 (NA = 0.1)
are both used to image the ion. The imaging system of obj 1 has a
magnification of 30. The main function of obj 2 is to calibrate the
position of the depletion beam to overlap with the ion. The calibration
procedure is as follows. First, we use a laser beam focused by obj
2 to simulate the trajectory of the ion's fluorescence. We adjust
obj 2's lateral and focal position to ensure that the simulation beam's
image coincides with the ion's image on CCD 1. Second, we use CCD
2 to locate the ion's position. We then fine-tune the direction of
the depletion beam to make their images overlap on CCD 2. Third, we
use the ion as an intensity probe to tune the depletion beam's position
more precisely. The depletion light and detection light are turned
on at the same time, resulting in Doppler cooling cycles. The intensity
of the depletion light is characterized in terms of photon counts
per second. We carefully tune the path of the depletion laser to ensure
that the ion at its maximum emission rate.

\section{Holographic beam reshaping with DMD}

\begin{figure}
\includegraphics[width=12cm]{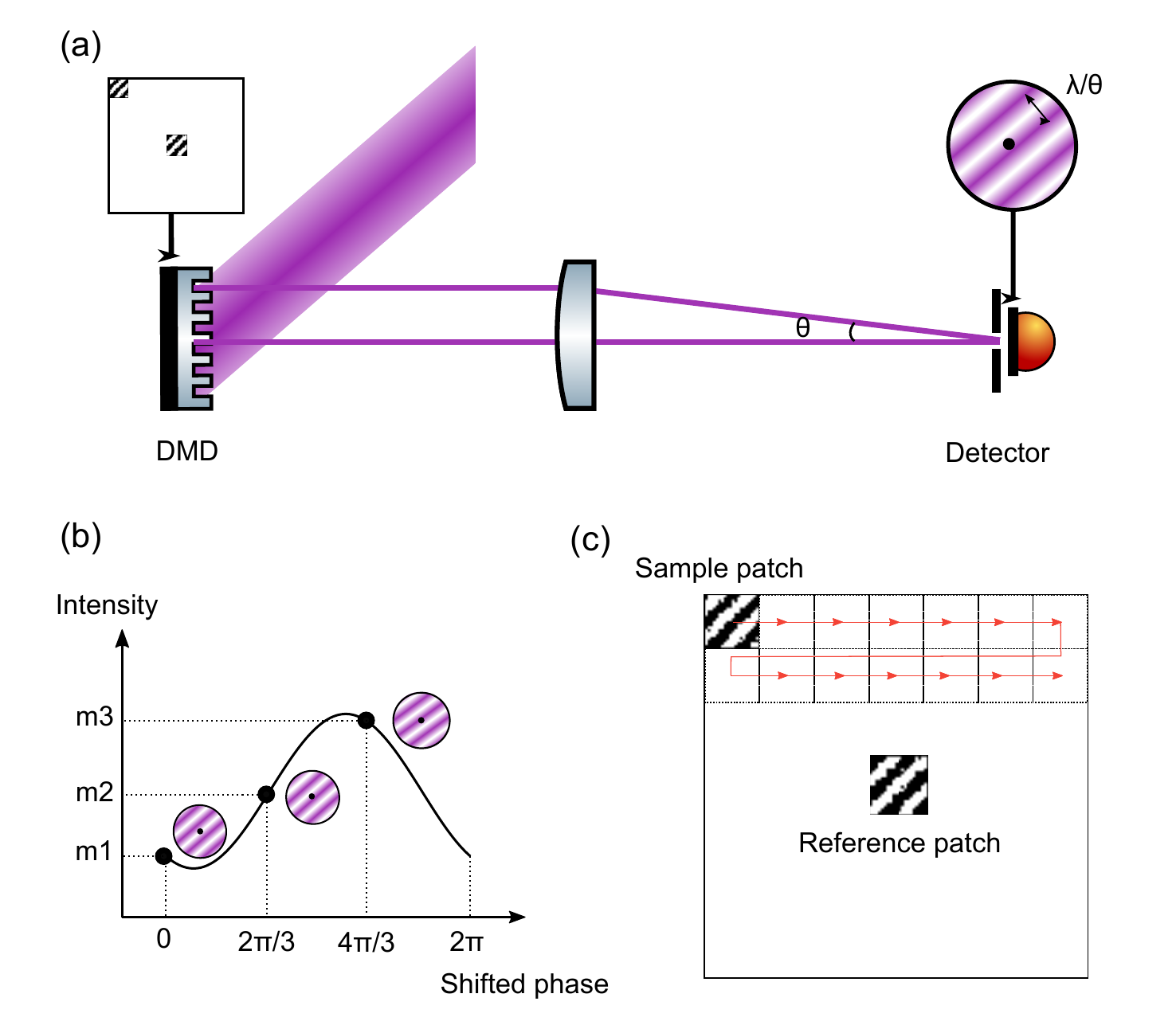} \caption{\label{fig:Fig6}Concept of aberration correction using a DMD. (a)
Interference of the beams diffracted by the sample patch and reference
patch. The DMD is illuminated by a collimated laser. The laser is
then focused by a lens. The first-order diffraction beams interfere
on the image plane. The detector measures the intensity at the central
position (black dot) of the interference pattern. The detector can
be an ion, CCD, or PMT. (b) Three-step phase-shift method to recover
the phase. The phase of the sample patch is shifted to 0, $2\pi/3$,
and $4\pi/3$, which results in three intensity values $m_{0}$, $m_{1}$,
and $m_{2}.$ The initial phase $\phi$ of the sample patch is calculated
from these. (c) Acquisition of the phase map by scanning the sample
patch while keeping the reference patch fixed. The default phase of
the reference patch is set to zero.}
\end{figure}

In ground-state depletion (GSD) super-resolution schemes, the first-order
Laguerre--Gaussian mode ($\mathrm{LG}_{0}^{1}$, $p=0$, $l=1$)
is often used as a donut beam. We use a holographic beam reshaping
method to produce the Laguerre--Gaussian mode. The main advantage
of this method lies in its ability to perform aberration sensing and
correction \citep{zupancic2016ultraprecise,Shih_2021}.

The holograph beam reshaping setup is illustrated in Fig.~\ref{fig:Fig5}
(gray box). The collimated depletion light is expanded 2.7 times by
L1 ($f_{1}=75$~mm) and L2 ($f_{2}=200$~mm). The 27~mm diameter
of the incident beam is large enough to cover the whole DMD area ($20.7\times11.7$~mm$^{2}$).
A holograph pattern on the DMD diffracts the beam into multiple orders.
The DMD area that we use is $660\times660$ pixels. The period of
the holograph grating is 10 pixels. The first-order beam is phase-modulated
and is selected by an iris. The diffraction efficiency is $\sim$5\%.
The piezo mirror (PI, S-330.2SH) is used to scan the donut spot on
the ion's plane in two dimensions; the scanning area ($50\times50~\upmu\text{m}^{2}$)
is large enough to cover the ion's trajectory. The scanning scale
between the tip/tilt angle and the displacement on the ion's plane
is 25~nm/$\upmu$rad, which is calibrated with the obj 2 imaging
system. The beam at the piezo mirror is conveyed to the entrance of
obj 1 through a 4-f system ($f=300$~mm). The entrance pupil diameter
of the spot is 7.1~mm.

To generate the donut beam, the first step is to measure the aberration
of the optical system via the DMD (ViALUX v9501). A DMD is a binary
spatial light modulator, which in this case is made of $1920\times1080$
aluminum micromirrors. Each micromirror has a width of 10.8~$\upmu$m.
It flips only $+12^{\circ}$ or $-12^{\circ}$ along a hinge corresponding
to on or off states. The main idea of phase calibration is to measure
the phase of a certain region (sample patch, $20\times20$ pixels)
relative to the phase of the center region (reference patch, $20\times20$
pixels) through a phase-shifting technique \citep{Liu_2015,Nov_k_2008}.
The procedure for aberration correction is as follows (see Fig.~\ref{fig:Fig6}).
First, we turn on the micromirrors of the sample patch and reference
patch. The phase of the reference patch is a global phase and so can
be regarded as zero. Two beams diffracted by holograph patterns will
interfere on the ion's plane. The ion serves as a detector of interference
pattern intensity. Shifting the phase of the sample patch will lead
to a shift of the interference pattern. The relative phase can thus
be measured via a three-step phase shift technique. We change the
sample patch while fixing the reference patch and repeat the above
steps until the whole phase map has been measured. The entire process
takes about 20~min.

\begin{figure}
\includegraphics[width=12cm]{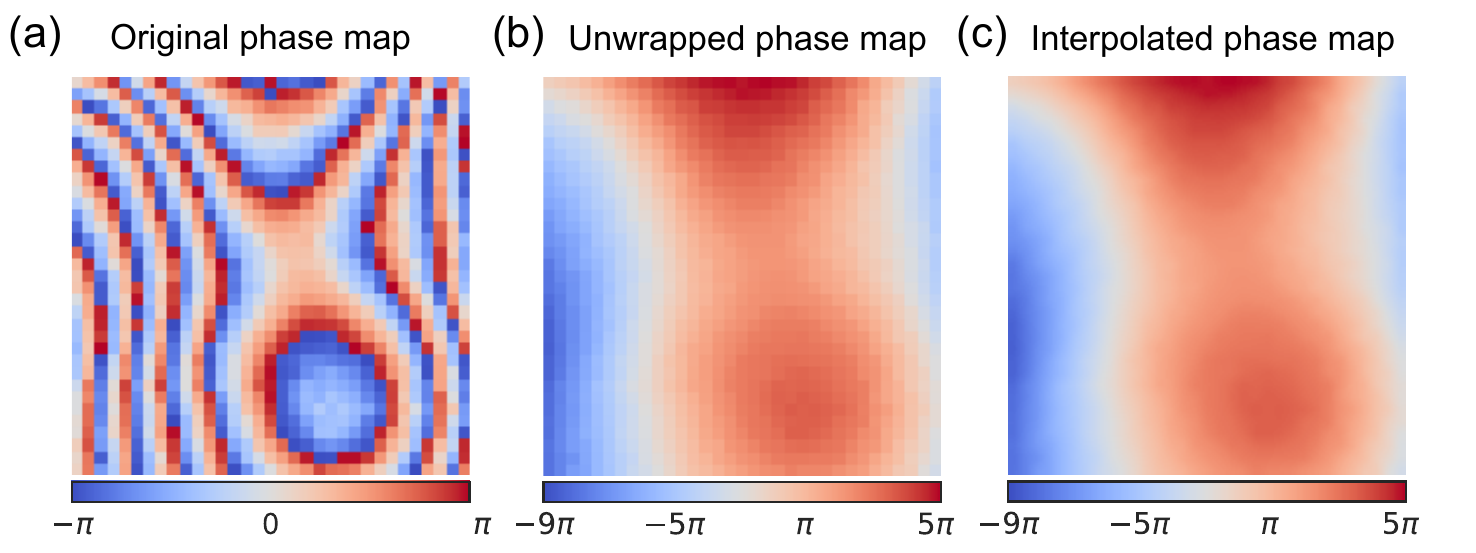} \caption{\label{fig:Fig7}Measured phase map. (a) Original phase map, whose
values lie in the range $[-\pi,\pi]$. (b) Unwrapped phase map. The
phase aberration ranges over $14\pi$, corresponding to $7\lambda$.
(c) Interpolation of (b) to each pixel.}
\end{figure}

The measured phase map is shown in Fig.~\ref{fig:Fig7}(a) and ranges
from $-\pi$ to $\pi$. It needs to be unwrapped to give a continuous
phase map. The unwrapped phase map {[}Fig.~\ref{fig:Fig7}(b){]}
ranges from $-9\pi$ to $5\pi$, indicating that the peak-to-valley
of aberration is $7\lambda$. The unwrapped phase is further interpolated
to each pixel to recover the tip/tilt terms of each patch. Thus, diffracted
beams focused by obj 1 can overlap well on the ion's plane, which
ensures higher accuracy of phase sensing. Finally, we obtain the smooth
phase map {[}see Fig.~\ref{fig:Fig7}(c){]}.

\begin{figure}
\includegraphics[width=12cm]{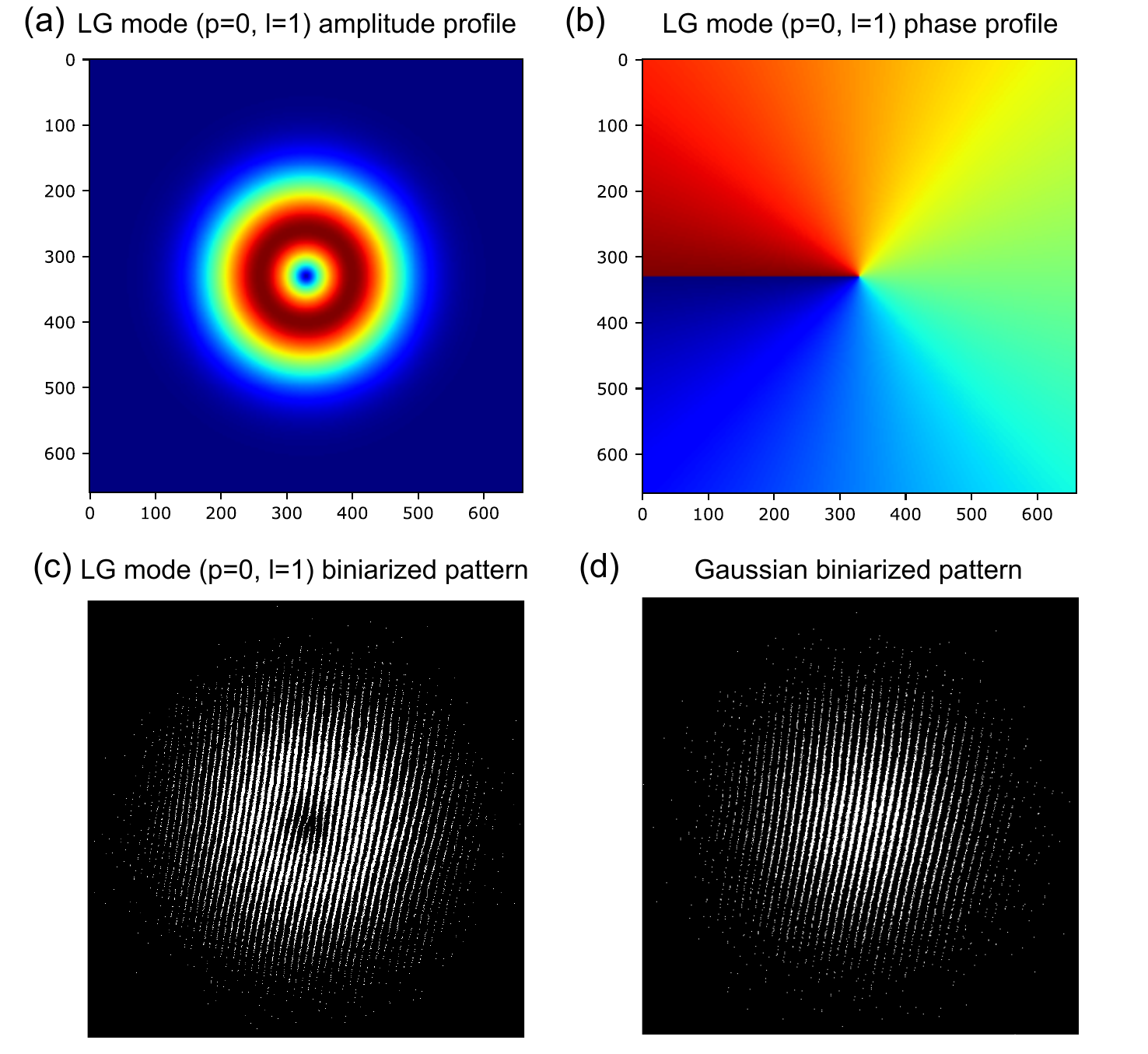} \caption{\label{fig:Fig8}Beam reshaping holographic patterns. (a) and(b) Phase
and amplitude profiles of the donut beam. (c) and (d) Holographic
patterns of donut modes and Gaussian modes on DMD.}
\end{figure}

The aberration measurement procedure is followed by beam reshaping.
To generate the Laguerre--Gaussian mode beam, phase and amplitude
modulation are implemented. The target phase is the sum of the aberration
phase maps in Fig.~\ref{fig:Fig7}(c) and the ideal Laguerre--Gaussian
phase map {[}Fig.~\ref{fig:Fig8}(b){]}. For amplitude modulation,
the ratio of ``on'' states in the computer-generated holograph pattern
is proportional to the amplitude profile of the Laguerre--Gaussian
mode {[}Fig.~\ref{fig:Fig8}(a){]}. Finally, the phase map and amplitude
map are binarized into a holograph pattern using the Lee method \citep{Lee_1979}.
The pattern is then projected on the DMD. With this method, a Gaussian
beam with high mode purity can also be achieved.

\begin{figure}
\includegraphics[width=12cm]{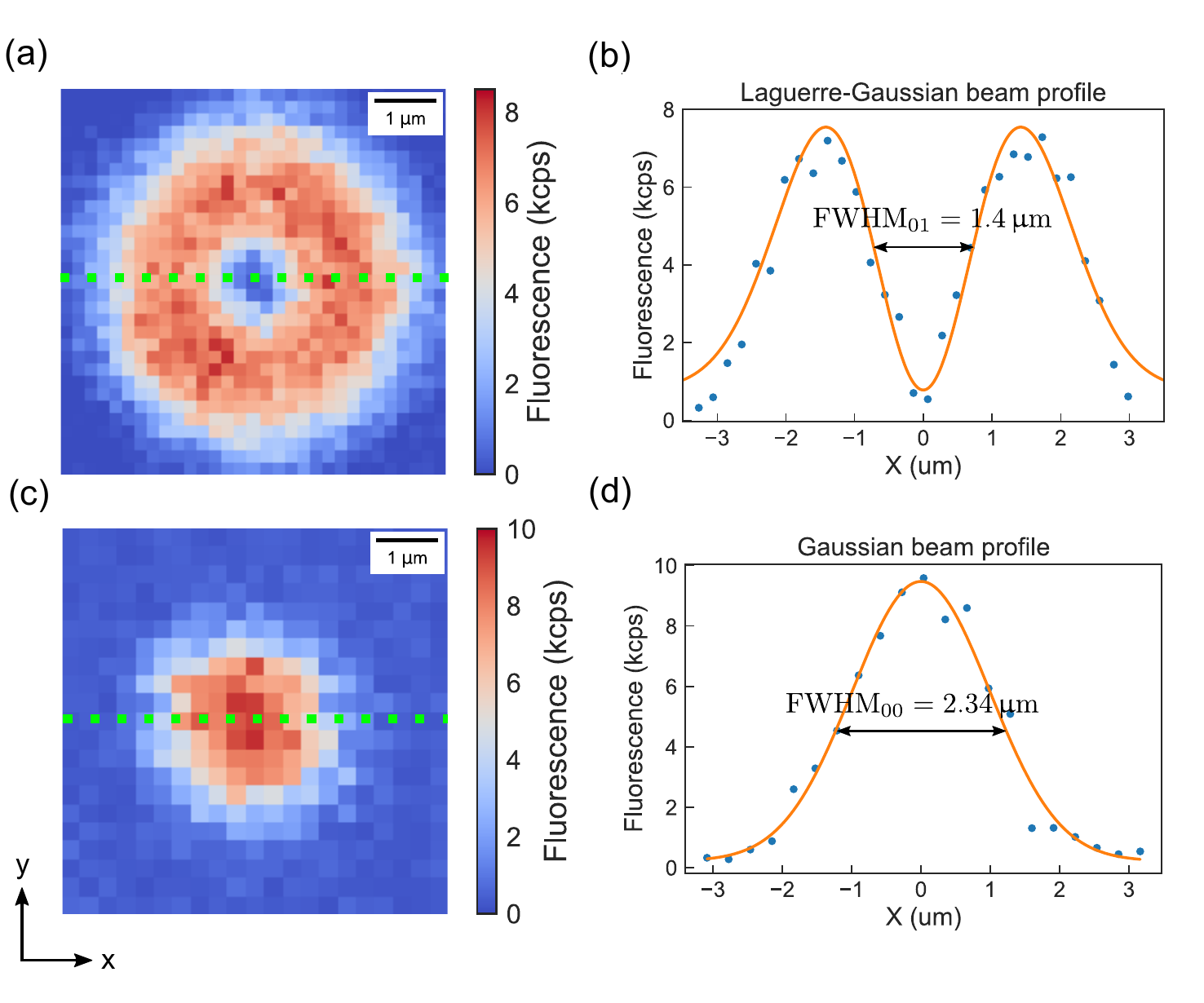}\caption{\label{fig:Fig9} Reshaped Gaussian and donut spots. The intensity
is represented by the number of fluorescence photons measured per
second (kcps = kilocounts per second). Background noise has been eliminated.
(a) and (b) Donut spot ($\mathrm{LG}_{0}^{1}$ mode). The minimum
intensity in the center is 3.8\% of the maximum intensity. The FWHM
of the central valley is 1.4 $\upmu$m. (c) and (d) Gaussian beam
with FWHM 2.34~$\upmu$m. The data for the curves in (b) and (d)
is extracted from the green dashed lines in (a) and (c). }
\end{figure}

The piezo mirror is scanned in two dimensions to measure the intensity
profiles of the donut spot and Gaussian spot. The power of the depletion
light here is 125~pW. The results are shown in Fig.~\ref{fig:Fig9}.
The FWHMs of the reshaped Gaussian and donut spots are 2.34~$\upmu$m
and 1.4~$\upmu$m, respectively. The residual intensity of the central
dark area is 3.8\% of the maximum intensity of the donut spot, which
verifies the high mode purity of the donut spot.

\section{Sequence of super-resolved imaging }

\begin{figure}
\includegraphics[width=12cm]{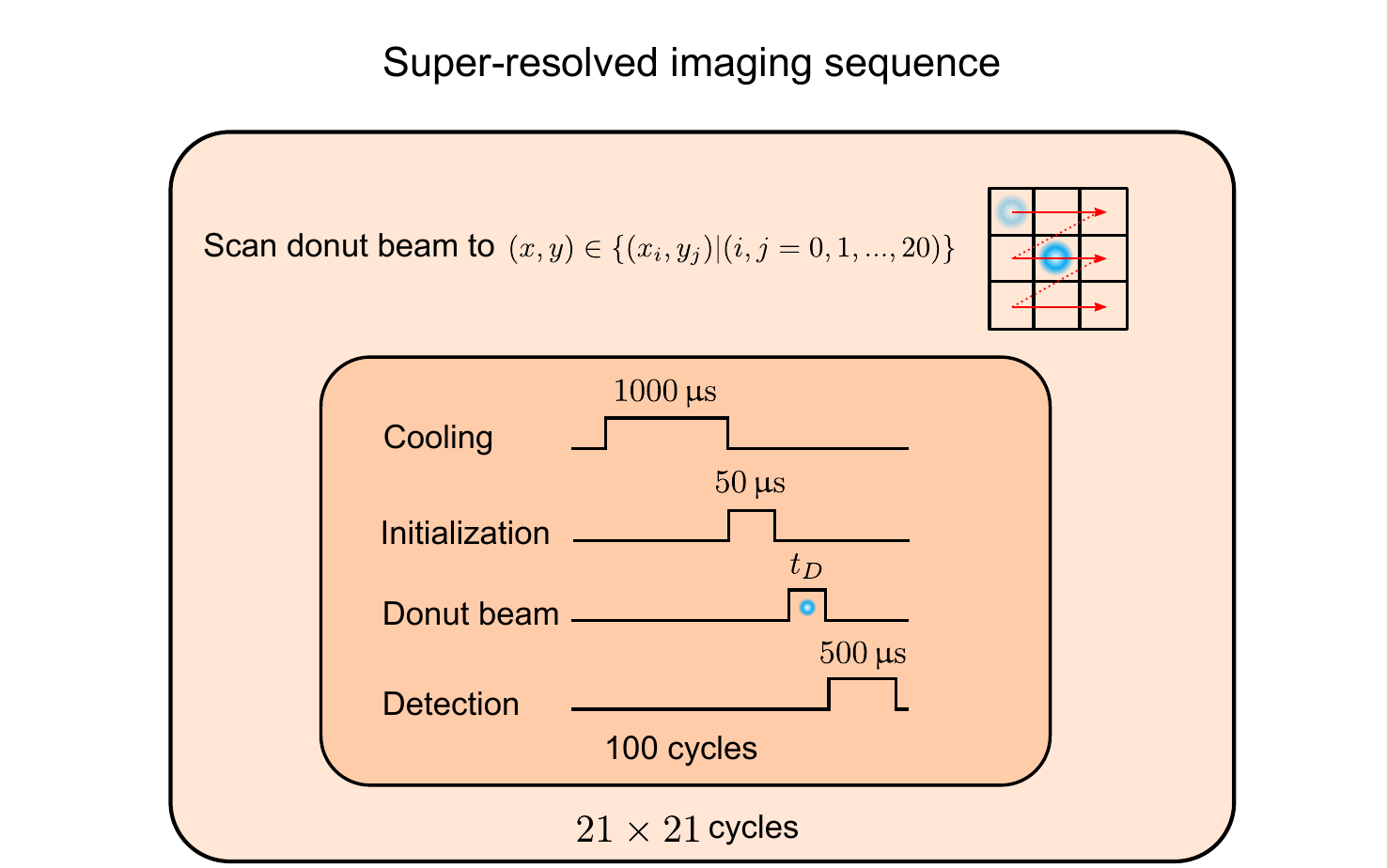} \caption{\label{fig:Fig10}Sequence of super-resolved imaging. It comprises
two layers of nested loops. At each position of the donut spot, the
inner layer of the sequence is cycled 100 times. }
\end{figure}

The sequence of super-resolved imaging is illustrated in Fig.~\ref{fig:Fig10}.
The whole sequence can be divided into two layers of nested loops.
In the outer layer, the piezo mirror is varied to scan the donut spot
over a series of positions $\{(x_{i},y_{j})\mid i,j=0,1,2,\dots,N\}$,
where $N=20$ in this work. At each position $(x_{i},y_{j})$, we
run the sequence in the second layer 100 times. The ion is first Doppler-cooled
for 1~ms. Then, the global pumping beam initializes the ion into
a dark state within 50~$\upmu$s. Next, the pulsed donut spot depopulates
the ion into a bright state. Finally, the ion's state is read out
by the detection laser. This inner layer cycles 100 times to produce
the dark-state probability. The position is then changed to the next
point, and the inner sequences are repeated.

In our current study, we just used a global initialization beam to
initialize the ion, since there are no other ions around it. This
is a simplified scheme of GSD. In the case of multiple atoms with
small spacing, to resolve a cluster of atoms with subdiffraction resolution,
instead of the global Gaussian beam, it is necessary to use a small
Gaussian spot, as in Fig.~2(a) of the main paper, to initialize the
emitter into the ground state locally, in the standard GSD microscopy
scheme. It is easy to achieve this requirement with our setup by projecting
Gaussian and donut holographic patterns on the DMD in sequence. The
lasers for initialization and depletion (turning on and off are controlled
by AOM 3 and AOM 4, respectively, in Fig.~\ref{fig:Fig3}) are combined
and guided to the DMD beam reshaping setup. In the initialization
process, we load the holograph pattern for a Gaussian spot on the
DMD and then switch AOM 3 on; after switching AOM 3 off to finish
the initialization, we then load the holograph pattern for the Gaussian
spot on the DMD. In the depletion process, we switch AOM 4 on for
the duration time for depletion. The holograph patterns have been
pre-programmed into the FPGA board of the DMD, and so we can rapidly
switching these patterns by an electrical trigger. The maximum switching
rate is 17\,857~Hz, which corresponds to a switch time of 6~$\upmu$s.
Because turning the lasers on and off is controlled by AOMs, the depletion
process can still be as fast as nanoseconds, and only the holograph
pattern switching process causes a time delay. To shorten the holograph
pattern switching time, we can use two DMD beam reshaping setups to
compensate and generate the Gaussian and donut spots individually.

\section{Motion frequency stabilization }

\begin{figure}
\includegraphics[width=12cm]{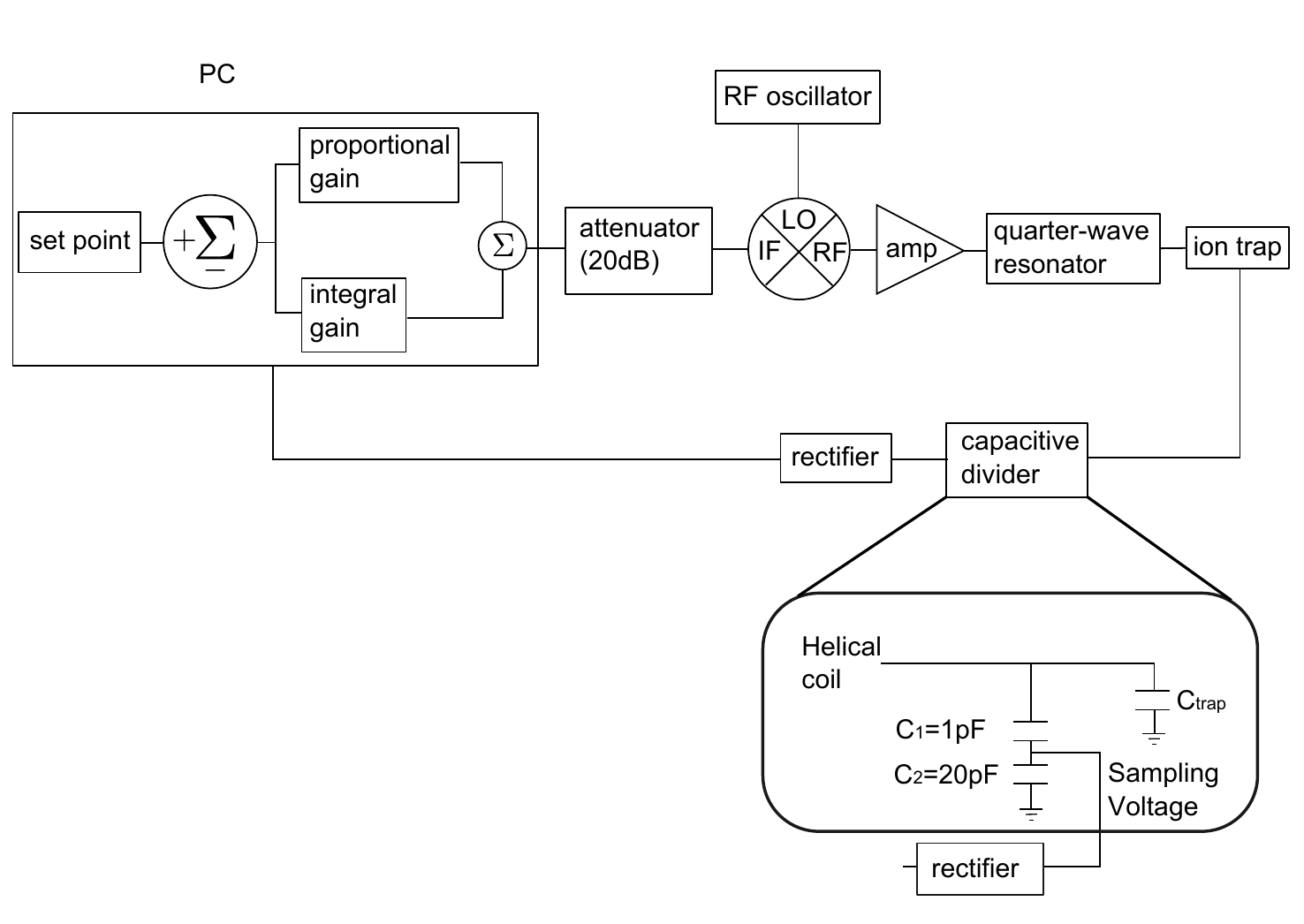} \caption{\label{fig:Fig11}Scheme of motion frequency feedback.}
\end{figure}

The ion's frequencies of motion on the $x$, $y$, $z$ axes are 804~kHz,
1.36~MHz and 847~kHz respectively, where the $y$ axis is defined
as lying along the central RF electrodes. To resolve the ion's dynamics,
we drive its motion on the $y$ axis with a resonant RF voltage. Therefore,
its motion frequency is required to remain stable under driving. To
stabilize this frequency, active feedback is designed and implemented
\citep{Johnson_2016}. The motion frequency is proportional to the
trapping voltage divided by the RF frequency. The drift of the trapping
voltage due to thermal fluctuations and amplifier noise is the main
reason for destabilization of the motion frequency. To stabilize this
frequency, feedback to the voltage on the trap electrodes is implemented.
The scheme is illustrated in Fig.~\ref{fig:Fig11}. A high voltage
is first sampled with a capacitor divider network comprising a 0.2~pF
and a 20~pF capacitor. The sampling AC voltage is 1\% of the trapping
voltage. This sampling voltage then passes through a rectifier and
is transformed into a DC voltage. This signal is acquired with an
analog-to-digital converter (ADC). The ADC that we use is a digital
multimeter (Rigol DM3068). It has a high accuracy of 6.5 bits and
a minimum resolved voltage of 1~$\upmu$V. The digital signal is
transmitted to a computer, which serves as a servo controller. It
runs a proportional--integral--derivative (PID) algorithm with a
set point 0.39~V and outputs a DC signal through a 24-bit digital-to-analog
converter (DAC). The signal is attenuated by 20~dB to make the output
voltage large enough. Thus, we can make the greatest use of the high
accuracy of the DAC. The attenuated signal is mixed with a local oscillator
(LO) signal generated by an RF source (22~MHz, $-25$~dBm). The
output signal is then amplified by 51~dB and finally transmitted
into the trap via a quarter-wave resonator.

\begin{figure}
\centering{}\includegraphics[width=8cm]{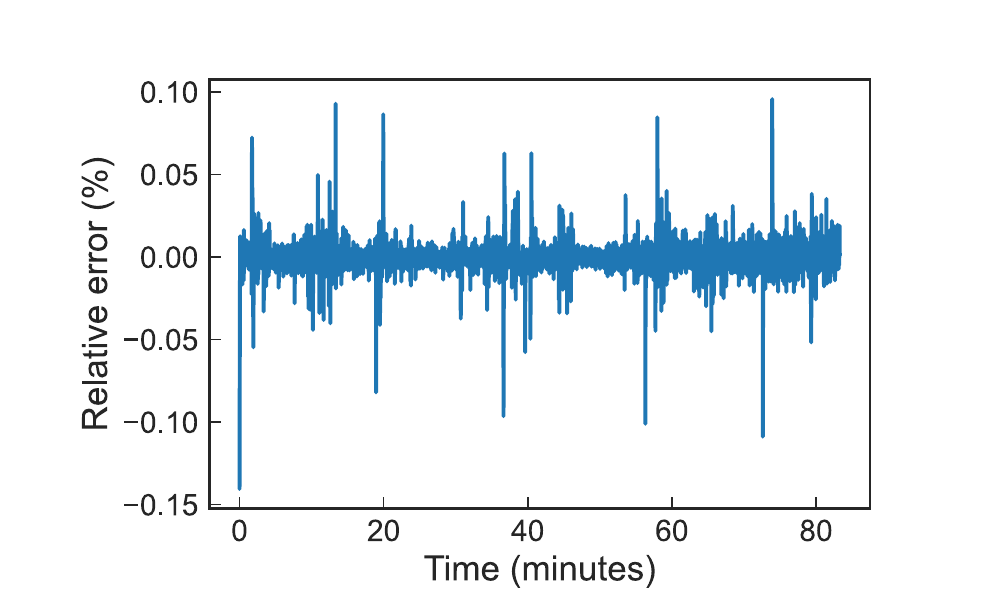} \caption{\label{fig:Fig12}Drift of the sampling voltage. The relative voltage
error $\delta V/V$ is defined as the difference between output voltage
and the set point divided by the set point.}
\end{figure}

The result is shown in Fig.~\ref{fig:Fig12}. The standard error
of the relative voltage $\delta V/V$ is suppressed to within 88~ppm.
The occasional spikes may result from the low feedback speed (2~Hz),
which is limited by the sampling rate of the ADC. The drift of the
motion frequency fluctuates within 0.5~kHz within an hour. In the
two-dimensional (2D) motion detection scheme, the RF driving time
is 73.5~$\upmu$s, which means that the fluctuation of the $y$-mode
motion frequency should be far less than 7~kHz. This feedback system
meets our requirements.

\section{Sequence of 2D motion detection}

The super-resolved technique is applied to resolve the ion's dynamics
in two dimensions. Here we drive the ion's $y$-mode motion with a
resonant RF signal (1.36~MHz).

\begin{figure}
\centering{}\includegraphics[width=12cm]{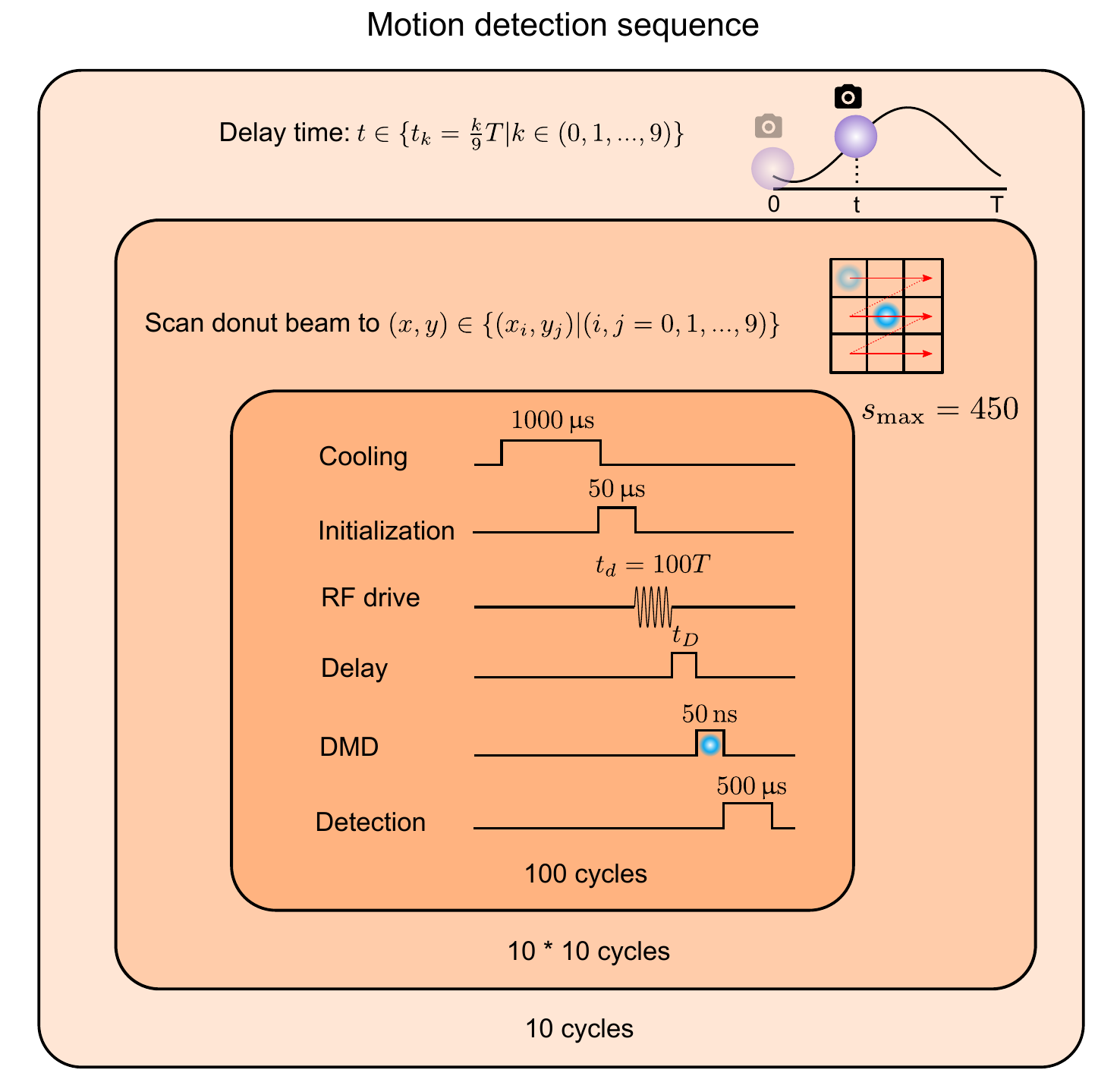} \caption{\label{fig:Fig13}Motion detection sequence. It comprises three layers
of nested loops. In the first layer, the delay time $t$ is set to
a series of values, meaning that the ion is observed at different
motion phases. In the second layer, the piezo mirror's angle is set
to scan the donut spot. In the third layer, the ion is driven by a
pulsed RF signal, and the depletion beam then takes a fast snapshot
of the moving ion. }
\end{figure}

The sequence is as shown in Fig.~\ref{fig:Fig13}. It is divided
into three layers of nested loops. In the first layer, we set the
delay time $t$ to a series of values to observe the ion's motion
at different oscillation phases. We sample 10 points within a motion
period ($T=735$~ns). In the second layer, similar to the super-resolved
sequence, we scan the donut spot to a series of positions $(x_{i},y_{j})$.
The scanning area is $1.25\times1.25~\upmu$m$^{2}$, with $10\times10$
points. In the last layer, we drive the ion for 100 periods with an
RF signal. We then stop driving and wait for a time $t$ followed
by a 50~ns depletion process with the donut spot. $I_{\mathrm{max}}/I_{\mathrm{sat}}$
is fixed at 450. This third layer is cycled 100 times to obtain the
dark-state probability.

\section{Depletion process simulation and control}

\begin{figure}
\includegraphics[width=12cm]{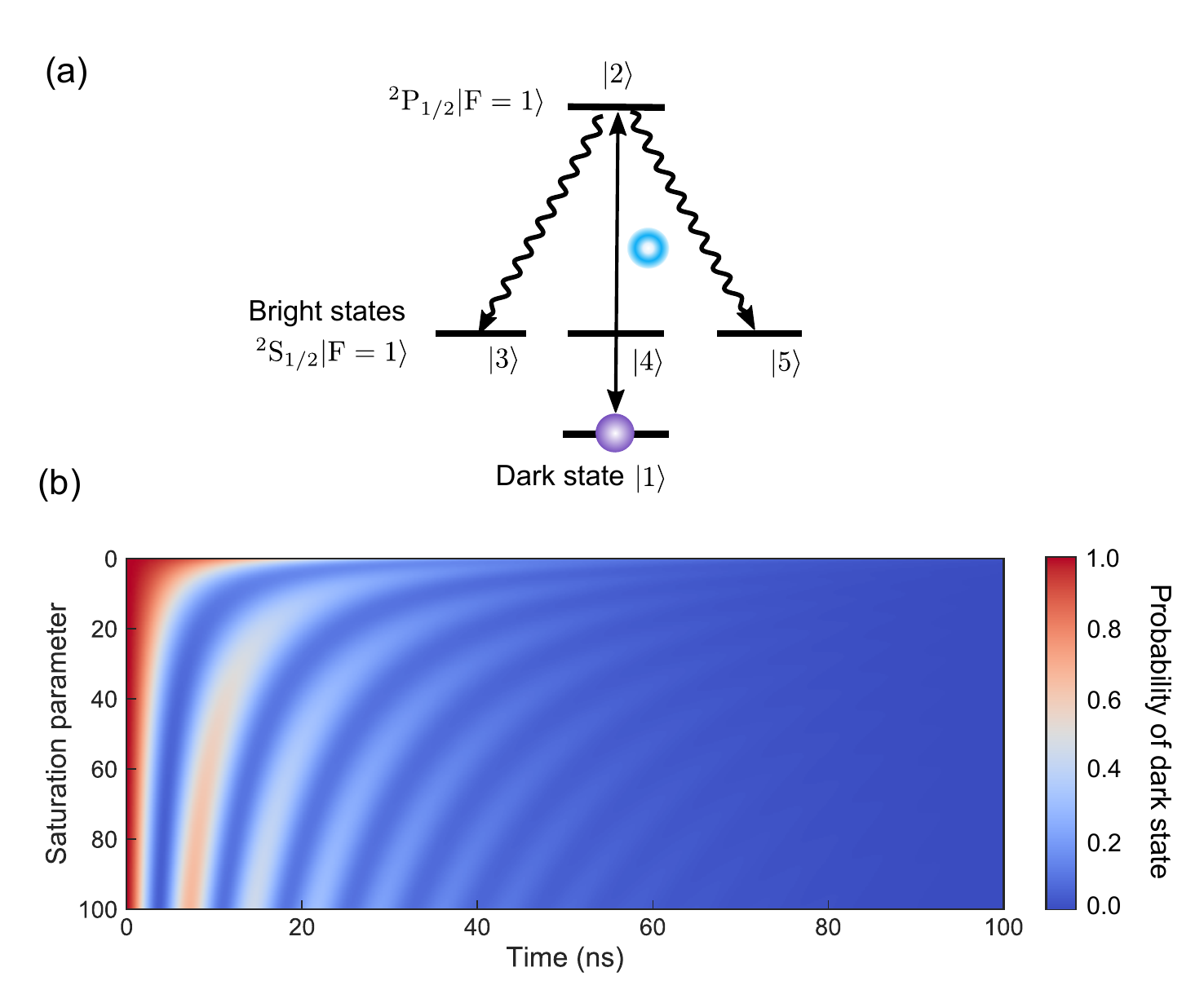} \caption{\label{fig:Fig14} Numerical simulation of depletion process under
different intensities. (a) Model of three-level system under a pulsed
depletion beam. (b) Probability of dark state vs time with different
intensities. The intensity is characterized by the saturation parameter.
As the saturation parameter increases, the speed with which the probability
approaches zero becomes much slower. }
\end{figure}

The depletion interaction can be modeled by an optical pumping process,
as shown in Fig.~\ref{fig:Fig14}(a). The probability evolution of
the dark state is described by the master equation 
\[
\partial_{t}\rho(t)=-\frac{i}{\hbar}[H,\rho(t)]+\sum_{n}\tfrac{1}{2}[2C_{n}\rho(t)C_{n}^{+}-\rho(t)C_{n}^{+}C_{n}-C_{n}^{+}C_{n}\rho(t)],
\]
where $H=\frac{1}{2}\Omega|1\rangle\langle2|+\mathrm{H.c.}$, $C_{1}=-\sqrt{\frac{1}{3}\Gamma}\,|1\rangle\langle2|$,
$C_{2}=\sqrt{\frac{1}{3}\Gamma}\,|3\rangle\langle2|$, $C_{3}=\sqrt{0\times\Gamma}\,|4\rangle\langle2|$,
and $C_{5}=-\sqrt{\frac{1}{3}\Gamma}\,|5\rangle\langle2|$ are the
collapse operators representing the decays of the excited state $|2\rangle$
to $|1\rangle$, $|3\rangle$, $|4\rangle$, and $|5\rangle$. $\Gamma=2\pi\times19.6$~MHz
is the spontaneous emission rate of $|2\rangle$. We simulate the
probability evolution of the dark state with QuTiP \citep{johansson2013qutip2}.
We use the saturation parameter $s$ to characterize the light intensity:
\begin{equation}
s=\frac{I}{I_{\mathrm{sat}}}=\frac{2\Omega^{2}}{\Gamma^{2}},
\end{equation}
where $I_{\mathrm{sat}}$ is the saturation intensity and $\Omega$
is the Rabi frequency.

The result is shown in Fig.~\ref{fig:Fig14}. For $s=100$, it takes
50~ns for the probability of the dark state to drop below 0.05. The
depletion time will not decrease linearly, even for larger values
of the saturation parameter. Considering that too short a donut pulse
would lead to inefficient depopulation of the dark state, we set the
depletion time to 50~ns. This pulse duration is far shorter than
the motion period (735~ns), which is sufficient to resolve the ion's
dynamics.

\begin{figure}
\includegraphics[width=12cm]{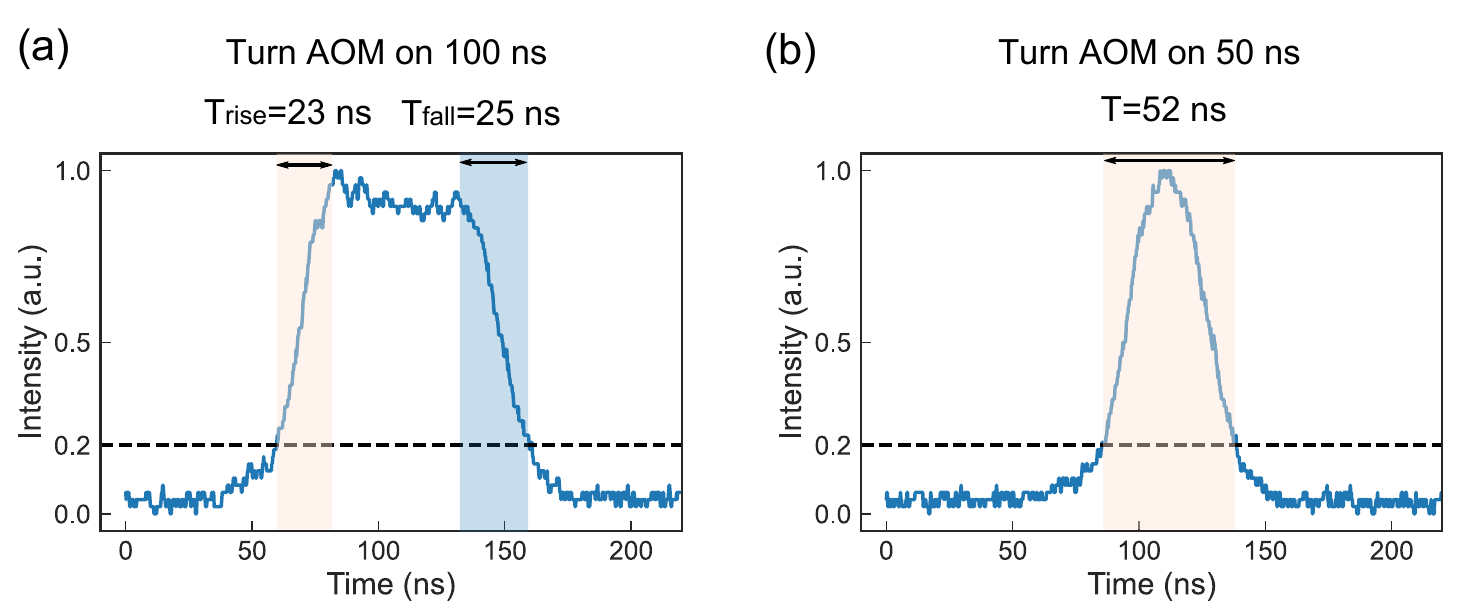} \caption{\label{fig:Fig15}Pulse shapes of the AOM. (a) Pulse duration $t_{D}=100$~ns:
the rise time is 23~ns and the fall time is 25~ns. (b) Pulse duration
$t_{D}=50$~ns: the full pulse duration measured is 52~ns.}
\end{figure}

In an experiment, the pulse duration is also limited by the rise time
of the AOMs, as illustrated in Fig.~\ref{fig:Fig15}. We observed
a rise time of 23~ns and a fall time of 25~ns from 20\% to 100\%
intensity. The modulation efficiency will drop dramatically when the
pulse time decreases below 50~ns.

\begin{figure}
\includegraphics[width=12cm]{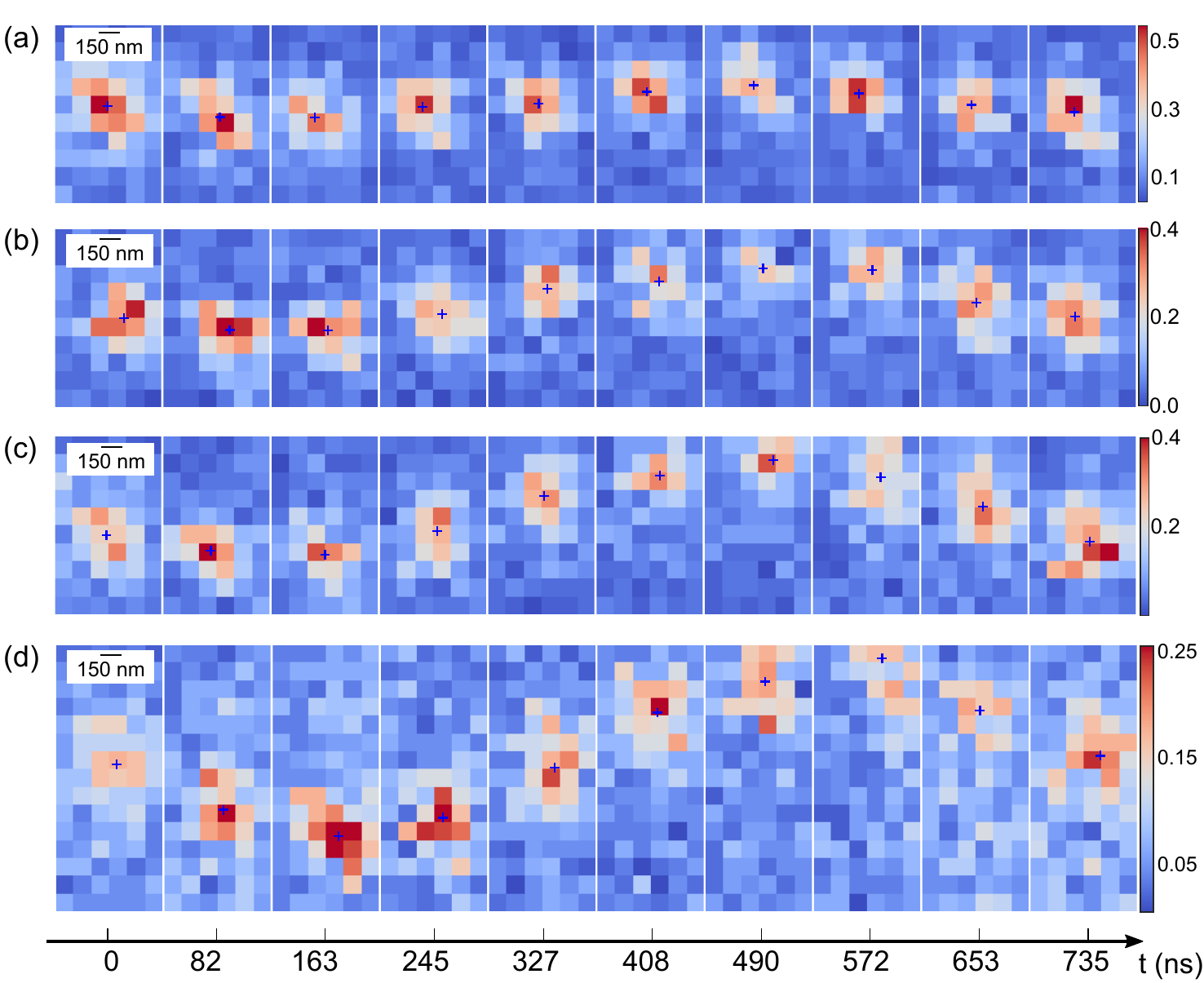} \caption{\label{fig:Fig16}Snapshots of ion's motion in a motion period under
driving voltages $V_{\mathrm{pp}}=38~\upmu$V (a), 82~$\upmu$V (b),
122~$\upmu$V (c), and 216~$\upmu$V (d).}
\end{figure}

Figure~\ref{fig:Fig16} depicts the ion's trajectory of motion under
driving voltages of $V_{\mathrm{pp}}=38$, 82~$\upmu$V, 122~$\upmu$V,
and 216~$\upmu$V. The ion's position from 2D Gaussian fitting is
indicated by a blue cross.

\begin{figure}
\includegraphics[width=12cm]{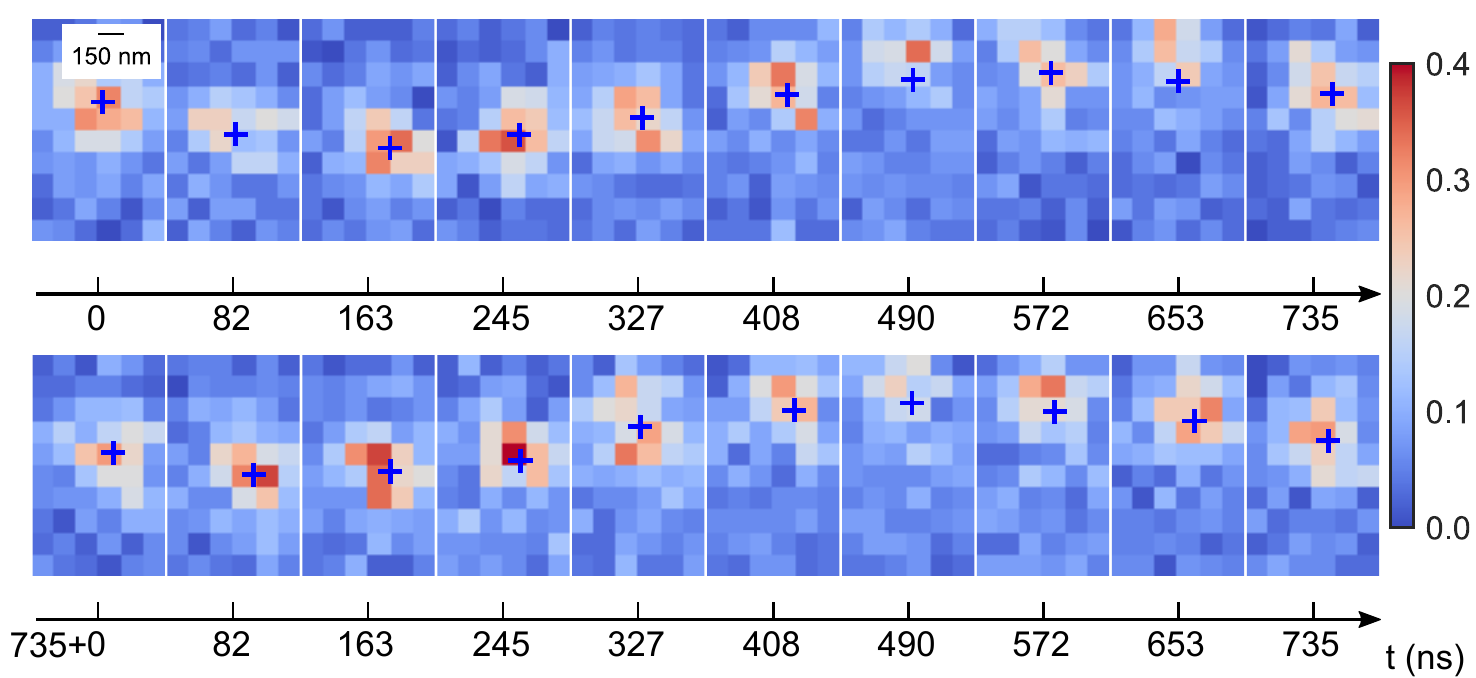} \caption{\label{fig:Fig17}Ion's trajectory of motion in two motion periods
under a driving voltage $V_{\mathrm{pp}}=216~\upmu$V. }
\end{figure}

We also measured the ion's motion in two periods (Fig.~\ref{fig:Fig17})
The motion period is fitted as $723.2\pm9.9$~ns. The relative error
is 1.6\%.

\section{Imaging resolution}

In this section, we will derive a formula for the resolution. The
ion is optically pumped to the bright state by a donut spot with duration
$t_{D}$. In this regime, the duration is much longer than the natural
lifetime ($t_{D}\gg1/\Gamma$). The dark-state probability $p_{0}$
can be modeled simply with the state rate equation 
\begin{equation}
\frac{dp_{0}}{dt}=-\beta\sigma\frac{I(x)}{\hbar\omega_{0}}p_{0},\label{eq:rate_equation}
\end{equation}
where $\beta=2/3$ is the branching ratio for the decay to bright
states, $\sigma$ is the cross-section, $\omega_{0}$ is the angular
frequency of the transition, and $I(x)$ is the intensity profile
of the donut spot in terms of position $x$. For generality, we use
the normalized intensity $s(x)=I(x)/I_{\mathrm{sat}}$, where $I_{\mathrm{sat}}=\hbar\omega_{0}^{3}\Gamma/12\pi c^{3}=510$~W/m$^{2}$
is the saturation intensity. The solution of Eq.~\eqref{eq:rate_equation}
is given by 
\begin{equation}
p_{0}(x)=\exp\!\left[-\frac{\beta\sigma}{\hbar\omega_{0}}I_{\mathrm{sat}}s(x)t_{D}\right],\label{eq:dark_state_prob}
\end{equation}
where $k=(\beta\sigma/\hbar\omega_{0})I_{\mathrm{sat}}$. The depletion
rate from dark state to bright state is $R_{x}\triangleq ks(x)$ and
is proportional to the intensity. Therefore, Eq.~\eqref{eq:dark_state_prob}
can be written as 
\begin{equation}
p_{0}(x)=\exp[-ks(x)t_{D}].
\end{equation}

The normalized intensity profile of the donut beam ($\mathrm{LG}_{0}^{1}$
beam) is 
\begin{equation}
s(x)=s_{0}+s_{\mathrm{max}}(x/r_{0})^{2}\exp\!\left[1-\left(\frac{x}{r_{0}}\right)^{2}\right],
\end{equation}
where $r_{0}=\mathrm{FWHM}_{01}$, i.e., the full width at half maximum
of the donut center, $s_{0}$ is the normalized residual intensity
of the center, and $s_{\mathrm{max}}$ is the maximum intensity of
the donut spot. The extinction ratio $\mathrm{ER}=s_{\mathrm{max}}/s_{0}$
is defined to characterize the ratio of the residual intensity of
the center. In the low-intensity regime ($s_{\mathrm{max}}\ll1$),
the bright-state probability $p_{1}=1-p_{0}$ is proportional to the
depletion light intensity. Thus, the resolution $\Delta r$ is equal
to $\mathrm{FWHM}_{01}$. In the high-intensity regime ($s_{\mathrm{max}}\gg1$),
the nonlinear response of the ion to light makes the resolution $\Delta r\ll\mathrm{FWHM}_{01}$.
Expansion of $s(x)$ at $x=0$ results in a parabolic shape: 
\begin{equation}
s(x)\approx s_{0}+es_{\mathrm{max}}\!\left(\frac{x}{r_{0}}\right)^{2}.
\end{equation}
Therefore, the population of the dark state follows as 
\begin{align}
p_{0} & =\exp(-ks_{0}t_{D})\exp\!\left[-kes_{\mathrm{max}}t_{D}\!\left(\frac{x}{r_{0}}\right)^{2}\right]\nonumber \\[6pt]
 & =\exp\!\left(-\frac{ks_{\mathrm{max}}t_{D}}{\mathrm{ER}}\right)\exp\!\left[-kes_{\mathrm{max}}t_{D}\!\left(\frac{x}{r_{0}}\right)^{2}\right].
\end{align}

\noindent The first factor $p_{0}^{\mathrm{max}}\triangleq\exp(-ks_{\mathrm{max}}t_{D}/\mathrm{ER})$
characterizes $p_{0}$ at the center of the donut spot, which will
exponentially decline as $s_{0}$ increases. The resolution can be
obtained from the second factor under the condition that this factor
is equal to one-half, i.e., $\exp[-kes_{\mathrm{max}}t_{D}(x_{0}/r_{0})^{2}]=1/2$.
Thus, the resolution is 
\begin{equation}
\Delta r=2x_{0}=\sqrt{\frac{4\ln2}{eks_{\mathrm{max}}t_{D}}}\,r_{0}\approx\frac{\mathrm{FWHM}_{01}}{\sqrt{ks_{\mathrm{max}}t_{D}}}.\label{eq:resolution}
\end{equation}
Uniting cases of low and high intensity, we obtain the formula for
the resolution: 
\begin{equation}
\Delta r=\frac{\mathrm{FWHM}_{01}}{\sqrt{1+ks_{\mathrm{max}}t_{D}}},
\end{equation}
which is very similar to that in conventional STED microscopy \citep{RN1601}.

The resolution is limited mainly by the normalized residual intensity
$s_{0}$. When the signal-to-noise ratio SNR = 1, i.e., $p_{0}^{\mathrm{max}}$
is equal to the error in dark-state measurement $\Delta p$, we can
easily extract the resolution from $p_{0}$: 
\begin{equation}
\exp\!\left(-\frac{ks_{\mathrm{max}}t_{D}}{\mathrm{ER}}\right)=\Delta p,
\end{equation}
and so 
\begin{equation}
ks_{\mathrm{max}}t_{D}=-\mathrm{ER}\ln\Delta p.
\end{equation}
Therefore, the resolution limit according to Eq.~\eqref{eq:resolution}
is 
\begin{equation}
\Delta r_{\mathrm{limit}}=\frac{\mathrm{FWHM}_{01}}{\sqrt{-\mathrm{ER}\ln\Delta p}},\label{eq:reoslutions2}
\end{equation}
and so $\Delta r_{\mathrm{limit}}\propto\lambda/\mathrm{NA}\sqrt{\mathrm{ER}}.$
For our experiment, the green dots in Fig.~2(f) of the main paper
indicate that the measurement noise $\Delta p\approx0.05$, $\mathrm{ER}=1/3.8\%=26$,
and $\Delta r_{\mathrm{limit}}=159$~nm, which is in agreement with
our measured best resolution of 175~nm.

In regardless of the position drift of the experiment system, here
we can give a theoretical resolution estimation of our method. In
our current study, the aberration of the objective is large, from
Fig.~\ref{fig:Fig7} , we can find the phase difference between two
adjacent pixels can almost reach $\pi$ at the edge of the phase map,
which means the real phase on the pixel is distributed in a range,
not a fixed value, so the aberration on the pixel can not be compensated
with a fix value. The large aberration of the objective makes the
compensation failure at the edge of the pupil of the objective. If
an objective with lower aberration and larger NA is used, e.g., NA=0.6,
the extinction ratio (ER) of the donut spot can be improved from 26
to 1000 \citep{zupancic2016ultraprecise,Yan_2015,Yan_2019}. From
the deduced resolution relation in the supplementary Eq.~\ref{eq:reoslutions2},
the resolution can be further improved to 4.7~nm with the NA and
ER improvement. Furthermore, the fidelity of single-shot readout of
the spin $\Delta p$ can be also improved with the high NA objective,
from 5\% to less than 0.1\% \citep{noek_high_2013}, which will further
improve the resolution to 3.1~nm.

\begin{figure}
\includegraphics[width=12cm]{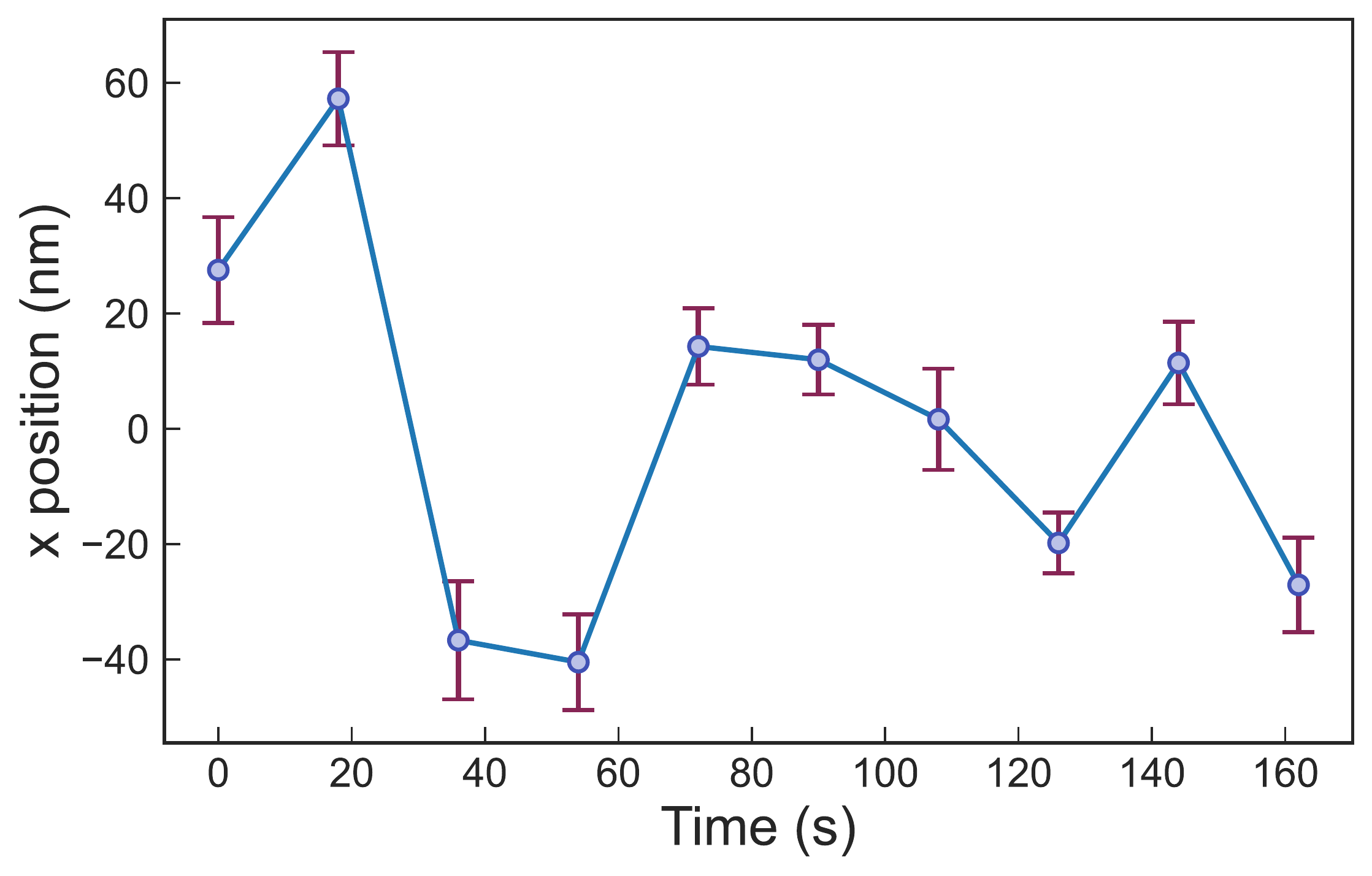} \caption{\label{fig:Fig18}Position drift of between the depletion spot and
the ion. By tracking the ion in the $x$-direction (the axis is not
driven) with different images in Fig. 3(c).}
\end{figure}
To realize the ultimate resolution, there will be lots of experimental
challenges. A typical problem is the stability of the optical system.
In our work, as the ion position displacement can be detected with
10~nm resolution, we can track the drift of the depletion light position
by imaging the ion at different time. As taking each image in Fig.
3(c) needs 16~s, (the ion area just occupies several pixels in an
image, so the imaging time on the ion area to get the displacement
is about 1s , the time to obtain the ion displacement is so short
that we can neglect the drift), we can get positions the ion in the
$x$-direction (the axis is not driven) at different experiment time,
which is shown in the Fig.~\ref{fig:Fig18}, it reflects the position
drift between the depletion spot and the ion. We can find the position
drift is around $\pm60$~nm in 160~s and an average drift speed
of 1.8~nm/s. It needed about 16~s to get the entire image in our
current experiment, the drift during this time is about 29nm. We must
note the estimated 29~nm is the drift after getting the entire image,
as the ion image just occupied several pixels in the entire image,
the drift during getting the ion image is much smaller in our experiment.
By using a high NA objective to perform the experiment, the experiment
time to get an image can be further compressed below 10~s, because
that the cooling time and detection time can be shortened from 1000us
to 100us \citep{noek_high_2013}, so the drift between images can
be further minimized. 

\bigskip{}
 Z.-H.Q. and J.-M.C. contributed equally to this work. 


%